\documentclass[submission, Phys]{SciPost}

\usepackage{amsmath,bm}
\usepackage{amsfonts}
\usepackage{float}
\usepackage{soul}
\usepackage{enumitem}
\usepackage{varwidth}
\usepackage{bigints}
\usepackage{tasks}
\usepackage{tikz}

\numberwithin{equation}{section} 
\newcommand{\cp}[1]{{\mathbb{CP}}^{#1}}
\newcommand{\td}[3]{|#1\;#2\;#3|}
\newcommand{\tp}[3]{(#1\;#2\;#3)}
\newcommand{\dd}[2]{|#1\;#2|}

\newcommand{\eso}[2]{s_{#1\;#2}}
\newcommand{\ese}[3]{s_{#1\;#2\;#3}}
\newcommand{\eq}[1]{Eq.~(\ref{#1})}

\newcommand{\one}{{\rm 1\kern -.9mm l}}

\newcommand{\be}{\begin{equation}}
\newcommand{\ee}{\end{equation}}

\allowdisplaybreaks

\begin{document}

\begin{center}{\Large \textbf{
Double Soft Theorem for Generalised Biadjoint Scalar Amplitudes}}\end{center}

\begin{center}
Md. Abhishek*,
Subramanya Hegde,
Dileep P. Jatkar,
Arnab Priya Saha
\end{center}

\begin{center}
Harish-Chandra Research Institute, Homi Bhabha National
 Institute (HBNI),\\
 Chhatnag Road, Jhunsi, Allahabad, India 211019
\\

* mdabhishek@hri.res.in
\end{center}

\begin{center}
\today
\end{center}

\section*{Abstract}
{\bf
We study double soft theorem for the generalised biadjoint
  scalar field theory whose amplitudes are computed in terms of
  punctures on $\mathbb{CP}^{k-1}$.  We find that whenever the double soft
  limit does not decouple into a product of single soft factors, the
  leading contributions to the double soft theorems come from the
  degenerate solutions, otherwise the non-degenerate solutions
  dominate.  Our analysis uses the regular solutions to the scattering
  equations.  Most of the results are presented for $k=3$ but we show
  how they generalise to arbitrary $k$.  We have explicit analytic
  results, for any $k$, in the case when soft external states are
  adjacent.
}

\vspace{10pt}
\noindent\rule{\textwidth}{1pt}
\tableofcontents\thispagestyle{fancy}
\noindent\rule{\textwidth}{1pt}
\vspace{10pt}

\section{Introduction and Summary}
\label{sec:intsum}

Our understanding of the scattering amplitudes has improved manifolds
in the last couple of decades.  That the formulae for amplitudes
simplify significantly if we use the spinor helicity formalism dates
back to the mid-eighties
\cite{Parke:1986gb,Nair:1988bq,Berends:1989hf}.  However, recent
interest in this direction came from the twistor formulation of the
${\cal N}=4$ super-Yang-Mills
theory\cite{Witten:2003nn,Roiban:2004yf}, leading to the BCFW
formulation of the scattering amplitudes\cite{Cachazo:2004kj,
  Britto:2005fq,ArkaniHamed:2008yf}.  The representation of the
${\cal N}=4$ super-Yang-Mills theory in terms of Grassmannians
\cite{Arkani-Hamed:2016byb,Arkani-Hamed:2017tmz,Arkani-Hamed:2017mur}
as well as polytope realisation of the scattering
amplitudes\cite{Arkani-Hamed:2016byb,He:2018pue,Salvatori:2018aha,
  Arkani-Hamed:2018rsk,Banerjee:2018tun,Raman:2019utu,Aneesh:2019ddi,
  Kalyanapuram:2019nnf,Aneesh:2019cvt,Salvatori:2019phs,
  Arkani-Hamed:2019vag,He:2020ray,Kalyanapuram:2020vil,Jagadale:2020qfa},
and the Cachazo-He-Yuan(CHY) formulation of the scattering amplitudes
\cite{Cachazo:2013iaa,Cachazo:2013gna,Cachazo:2013hca,Cachazo:2013iea,
  He:2015yua,Cachazo:2015aol} gave further impetus to unraveling their
underlying structure.

The space of Mandelstam invariants of $n$ massless particles is
isomorphic to the moduli space of $\cp{1}$ with $n$ marked points, the
CHY formulation therefore naturally involves the punctured Riemann
surfaces and in particular, for tree level $n$-point amplitudes they
were written in terms of the integral over the moduli space of
$n$-punctured sphere $\cp{1}$.  The CHY amplitudes were generalised in
different ways, which led to developments such as the ambitwistor
string theory
\cite{Mason:2013sva,Geyer:2014fka,Casali:2015vta,Geyer:2015bja,
  Geyer:2015jch,Geyer:2016wjx,Geyer:2017ela,Geyer:2018xwu,
  Berkovits:2019bbx}, and the positive Grassmannian formulation
\cite{Arkani-Hamed:2016byb,Arkani-Hamed:2017tmz,Arkani-Hamed:2019mrd,
  Arkani-Hamed:2020cig,He:2020onr} of string theory and field theory
amplitudes.  Recently one of the promising generalisations involved
replacing $\cp{1}$ by $\cp{k-1}$
\cite{Cachazo:2018wvl,Cachazo:2019ngv}.  This generalisation, in some
sense, is straightforward from the Grassmannian point of view but is
not at all obvious from the field theory point of view.  In other
words, going from $Gr(2,n)$ to $Gr(k,n)$ seems like a natural thing to
attempt from the Grassmannian picture, but it is not clear what kind
of field theories for which this generalised formulation of scattering
amplitudes applies.  In fact, it was known that N$^{k-2}$MHV
amplitudes can be written in terms of $Gr(k,n)$ \cite{Franco:2014csa}
well before this generalization\cite{Cachazo:2018wvl,Cachazo:2019ngv}
was proposed, which indicates that this formulation could be useful in
unravelling the structure of loop amplitudes.  Nevertheless, the
generalization of the biadjoint scalar theory with $\cp{k-1}$
kinematic space begs for a field theory formulation, which presumably
would give a better insight and provide more physical methods for
dealing with these amplitudes.  There has been some progress in
understanding these amplitudes for $k>2$ in terms of planar arrays of
Feynman diagrams
\cite{Borges:2019csl,Cachazo:2019xjx,Guevara:2020lek}.

In this paper, we will study double soft limits of amplitudes in the
generalised biadjoint scalar field theory.  In
\cite{Sepulveda:2019vrz} it was shown that single soft limits of the
$(k,n)$ amplitudes admit identification with the $(2,k+2)$
amplitudes.  Whether such an identification can be generalised to
multiple soft limits is an interesting question to explore.  Here we
will take the first step in this direction by computing double soft
factors in the generalised biadjoint scalar theories.  Since $(k,k+1)$
amplitudes are trivial by a gauge choice, the first non-trivial single
soft limit is applicable to $(k,k+2)$ amplitudes.  Naturally, the
first non-trivial case of soft limit with $m$ number of soft external
legs can be applied to $(k,k+m+1)$ amplitudes.  Using the Grassmannian
duality these amplitudes are related to $(m+1,k+m+1)$ amplitudes.  In
particular, for the double soft $(m=2)$ factors, we naturally expect
them to be related to $(3,k+3)$ amplitudes.  In addition note that as
the dimension of the moduli space for $(k,n)$ amplitudes is
$(k-1)(n-k-1)$, the multi soft limits mentioned above probe the
maximal codimension boundaries of the $m(k-1)$ dimensional moduli
space.

With the motivation given above, we will explore the structure of
double soft factors in arbitrary $(k,n)$ case.  We will, however, give
a more comprehensive account of our results for $k=3$ case, {\em
  i.e.}, when the amplitudes are described as punctures on $\cp{2}$,
and then generalise them to arbitrary $k$.  We will, however, not
explicitly use the positive Grassmannian formulation here.  Single and
multiple soft theorems in a variety of theories including gluons and
gravitons have been worked out using the CHY formalism \cite{Schwab,
  Afkhami, Zlotnikov, Kalousios, DoubleSoftPRD, VolovichZlotnikov,
  Saha:2016kjr, Saha:2017yqi, Chakrabarti:2017zmh}.  The biadjoint
scalar field theory has been the main arena for exploration, both in
the CHY formalism as well as its generalisation to higher $k$
\cite{Cachazo:2018wvl,Cachazo:2019ngv}.  By virtue of being a
biadjoint field, its amplitudes are parametrised in terms of two sets
of adjoint indices, we will denote them as $\alpha$ and $\beta$,
respectively.  The amplitudes can have independent color ordering with
respect to these two adjoint indices.  Throughout this work, we will
choose both $\alpha$ and $\beta$ to be the canonical ordering, that
denoted as $\mathtt{I}$, we will use this notation for the sake of
brevity.

The solutions to scattering equations for $k\ge 3$ can be categorised
into two types, regular and singular solutions, based on the behaviour
of kinematic determinants in the soft limit.  We will exclusively be
focusing on the regular solutions.  In addition, in this work we will
be interested in the leading contribution to the soft theorem and
hence we will pay attention to those configurations that give dominant
contribution to the single or double soft limit.  We find that, in the
double soft limit, those configurations that do not factorise into a
product of two single soft configurations have dominant contributions
coming from the degenerate solutions.  In all these cases, the
non-degenerate solutions lead to subleading contributions and hence,
in this paper, we do not take these cases into account.  The
non-decoupling double soft configurations for arbitrary $k$ occur when
two soft particles in the amplitude are such that separation between
them, for the canonical ordering, is not more than $k-2$.  For
example, for $k=2$, they must be adjacent external states, whereas for
$k=3$ they can be adjacent or the next to adjacent.  For any
configuration with index separation larger than this leads to the
double soft factor which is a product of two single soft factors.

The double soft limit contains two main cases, simultaneous double
soft limit and consecutive double soft limit.  In the simultaneous
double soft case, two external states are taken soft at an equal rate,
and in the latter case, one state becomes soft at a faster rate than
the other.  The leading contribution to the simultaneous double soft
limit comes from the degenerate solutions which have singularities
corresponding to collision of two soft punctures or corresponding to
collinear limit of two soft punctures with a hard puncture.  In the
latter case, two soft states scale differently with either
$\tau_1\ll \tau_2$ or $\tau_1\gg \tau_2$, where soft limit corresponds
to $\tau_i\to 0$ in a sequential manner.  We also establish that
simultaneous double soft limit can be arrived at by taking
$\tau_1=\tau_2$ limit of the consecutive limit.  In some sense, this is
a consistency check for our computation of the double soft limits.

For arbitrary $k$, the simultaneous double soft factor for the
adjacent soft external states is given by,
\begin{equation}
  \label{eq:mainresult}
  \mathtt{S}_{\text{DS}}^{(k)}  =
  \frac{1}{\sum\limits_{1\le a_{1}\cdots <a_{k-2}\le n-2}
    s_{a_{1}\cdots a_{k-2}\; n-1\; n}} \;\mathtt{S}^{(k-1)}
  \left(s_{a_{1}\cdots a_{k-2}\; m} \to s_{a_{1}\cdots a_{k-2}\; n-1\; n} \right)
  \; \mathtt{S}^{(k)}\ ,
\end{equation}
where $s_{a_1\cdots a_k}$ are generalised Mandelstam variables.  In
\eq{eq:mainresult}, we have taken the external states $n$ and $n-1$
soft, and the argument of the single soft factor $\mathtt{S}^{(k-1)}$
signifies that the soft label $m$ for $\mathtt{S}^{(k-1)}$ is replaced
by a composite label `$n\;n-1$'.  The single soft factor,
$\mathtt{S}^{(k)}$ is defined with a shifted propagator
$s_{a_{1}\cdots a_{k-1}\; n-1} + s_{a_{1}\cdots a_{k-1}\; n}$.  This
is the main result of this paper.  We also find, for any $k$, that the
leading contribution to the double soft factor scales as
$\tau^{-3(k-1)}$ in the $\tau\to 0$ limit.

We generalise our analysis to the next to adjacent soft external
states for $k=3$, where we encounter a high degree polynomial equation
to solve for the punctures in terms of the generalised Mandelstam
variables.  We, however, do not have generic explicit solutions to
this polynomial equation.  In the case of $k=3$, we show that the double
soft factor for the next to next to adjacent soft external states
factorises into a product of two $k=3$ single soft factors for each of
the soft external states.

The paper is organised as follows: Section \ref{sec:stk2} is more of a
review of the $k=2$ CHY formalism.  In this section we will study
single and double soft limits of $n$ point amplitudes in the biadjoint
scalar field theory.  We, however, will present the results for the
double soft limit of the amplitudes with non-adjacent soft punctures.
As in the literature, for an arbitrary $k$, we will use phrases like
punctures on $\cp{k-1}$, external states, and external particles
interchangeably.  In section \ref{singlesoft3n}, we will discuss the
single soft theorem for arbitrary $k$.  After setting up the notation,
we will consider single soft theorem in $k=3$ case and analyse
collision as well as collinear singularities.  We then generalise
these results to arbitrary $k$.  This section summarises the results
of \cite{Sepulveda:2019vrz}, but the method spelt out in this section
is useful for generalisation to the double soft limit.  Section
\ref{sec:dstg3n} contains a detailed study of the double soft theorem
for $k=3$, where we consider two external soft states to be adjacent.
Section \ref{sec:gen-k-adjacent} contains generalisation of the double
soft theorem for adjacent external states to arbitrary $k$.  In
section \ref{sec:nasl3a}, we revert to the $k=3$ case and study the
next to adjacent double soft limit of $n$ point amplitudes and in
\ref{sec:nnasl3}, we study the next to next to adjacent double soft
limit.  We conclude with section \ref{sec:disn}, which contains a
discussion on applications and possible extensions of these results.

\section{Soft Theorems for Biadjoint Scalar Field for $k=2$}
\label{sec:stk2}

In this section, we review the single and double soft limits of
biadjoint scalar amplitudes for $k=2$ in the CHY formalism.  As
mentioned in the introduction, we will take both $\alpha$ and $\beta$
to be canonically ordered, $\mathtt{I} = 1,2, \ldots n$.  Here we
shall consider soft limits in
$m_{n}^{(2)}\left(\mathtt{I}|\mathtt{I}\right)$. In the resulting
lower point amplitude canonical ordering would mean labels are
arranged in ascending order of magnitude after omitting the soft
particles.

\subsection{Single soft limit}
\label{Sec:single-soft-2}

We will begin with the familiar single soft limit of the $n$-point
amplitude of the biadjoint scalar field theory.  We will present the
computations in both homogeneous and inhomogeneous coordinates, which
helps us set up the notation for the rest of the paper.  In the
homogeneous coordinates, we can express the punctures on $\cp{1}$ by,
\begin{equation}
\sigma_{a} = \begin{pmatrix}
Z_a^1 \\ Z_a^2
\end{pmatrix}
= Z_a^1 \begin{pmatrix}
1 \\ x_a
\end{pmatrix}, \qquad x_a = \frac{Z_a^2}{Z_a^1}\ ,
\end{equation}
where $Z_a$ are homogeneous coordinates and $x_a$ are projective
coordinates defined in the coordinate patch where $Z_a^1$ is
non-vanishing.  It is convenient to introduce a potential
function\cite{Cachazo:2016ror},
\begin{equation}
  \label{eq:k2potfn}
  {\cal S}^{(2)} = \sum\limits_{b \ne a} \eso{a}{b}\log \dd{a}{b}\ ,
  \qquad \dd{a}{b} = \begin{vmatrix}
1 & 1 \\
x_a & x_b 
\end{vmatrix}\ ,
\end{equation}
whose extremisation gives the scattering equations, which in the
projective coordinates take the form,
\begin{equation}\label{scat-eqn-k2}
  E_a := \sum\limits_{b \ne a} \frac{\eso{a}{b}}{x_a - x_b} =
  \frac{\partial}{\partial x_a} \sum\limits_{b \ne a} \eso{a}{b}
  \log \dd{a}{b} = 0\ ,
\end{equation}
where $\eso{a}{b}$ are the Mandelstam variables. At this stage
although we can treat the Mandelstam variables in terms of specific
functions of momenta, e.g., $s_{a\; b} = 2 k_{a}\cdot k_{b}$, for
the purpose of generalisation to arbitrary $k$ we will keep them
generic without referring to explicit dependence on momenta.  We now define,
\begin{align}
  \label{eq:bell}
  E'_a
  := & \sum\limits_{b\ne a} \frac{\eso{a}{b}}{\dd{a}{b}}\dd{X}{b}
        \nonumber\\
  = & \sum\limits_{b\ne a} \frac{\eso{a}{b}}{\dd{a}{b}}(x_b- x_a+ x_a- x)
      \nonumber\\
  = & \dd{X}{a} \sum\limits_{b\ne a} \frac{\eso{a}{b}}{\dd{a}{b}} \nonumber\\
  = & - \dd{X}{a} E_a\ .
\end{align}
Here $X$ is an arbitrary reference vector on $\mathbb{CP}^{1}$ and we
will denote it by $X = \begin{pmatrix} 1 \\ x
\end{pmatrix}$. In the third equality in \eq{eq:bell} we have used the
condition of momentum conservation. It follows from the \eq{eq:bell}
that the delta function for $a$-th scattering equation can be
expressed as,
\begin{equation}
\delta(E_a) = - \dd{X}{a} \; \delta(E_a').
\end{equation}
In order to take the single soft limit, we need to choose soft momenta for
one of the external legs.  Without loss of generality, we will choose
$n$-th particle momentum to be soft.  Since the external momenta are
in one to one correspondence with the punctures on $\cp{1}$, we will
use the words momentum and punctures interchangeably. In the later
sections, where there is no clear description in terms of momenta, we
will only use the term punctures.  We will denote the soft puncture by
$\sigma$ in homogeneous coordinates.  In this coordinates the one-form
on $\mathbb{CP}^{1}$ can be written as,
\begin{eqnarray}
\left(\sigma\, d\sigma\right) & = & \begin{vmatrix}
Z^1 & dZ^1 \\
Z^2 & dZ^2
\end{vmatrix} \nonumber\\
& = & (Z^1)^2 dx_n, \qquad x_n = \frac{Z^2}{Z^1}.
\end{eqnarray}
Since the $n$-th external state has soft momentum, the $n$-particle
amplitude factorises into $n-1$-particle amplitude times the soft
factor.  We therefore denote,
$m_{n}^{(2)}\left(\mathtt{I}|\mathtt{I}\right) = \mathtt{S}^{(2)}\;
m_{n-1}\left(\mathtt{I}|\mathtt{I}\right)$, where the soft factor given by,
\begin{align}\label{chy-single-soft}
  \mathtt{S}^{(2)}
  = & -\oint\frac{(\sigma\; d\sigma)(X\,\sigma)}{\sum\limits_{b\ne n}
      \frac{\eso{n}{b}(X\,b)}{(\sigma\, b)}}\left[\frac{(n-1\, 1)}{
      \underline{(n-1\, \sigma)(\sigma\, 1)}}\right]^{2} \nonumber\\
  = & -\oint \frac{dx_{n}\left(x_{n} - x\right)}{\sum\limits_{b\ne n}
      \frac{s_{nb}\left(x_{b} - x\right)}{x_{b} - x_{n}}}\left[\frac{x_{1}
      - x_{n-1}}{\underline{\left(x_{n} - x_{n-1}\right)\left(x_{1} - x_{n}
      \right)}}\right]^{2} \nonumber\\   
 = & \frac{1}{\eso{n}{n-1}} + \frac{1}{\eso{n}{1}}.
\end{align}
In the last line we have used the residue theorem to evaluate
contributions coming from simple poles at $x_{n} = x_{n-1}$ and
$x_{n} = x_{1}$.

\subsection{Double soft limits}
\label{Sec:double-soft-2}

We will now study simultaneous soft limits with two external soft
momenta.  For the biadjoint scalars there are qualitatively two
different ways of taking simultaneous soft limits: in a given color
ordered arrangement, either two adjacent states are taken soft, or two
non-adjacent states are taken soft.  As we will see below in the
former case, the soft factor scales as $\tau^{-3}$, whereas in the
latter case, the scaling is $\tau^{-2}$.

\subsubsection{Adjacent soft limit}

As in the previous subsection, we will continue to take $k_n$, the
momentum of $n$-th particle soft.  In addition we will consider
$k_{n-1}$, the momentum of $(n-1)$-th particle, to be soft as well.
We implement these soft limits by scaling the soft momenta as
$k_n = \tau \hat{k}_n$ and $k_{n-1} = \tau \hat{k}_{n-1}$ and take
$\tau\to 0$.  As a result the Mandelstam variables scale in the
following manner, $s_{n-1\;a}=\tau\hat{s}_{n-1\;a}$ and
$s_{n\;a}=\tau\hat{s}_{n\;a}$, on the other hand
$s_{n-1\; n} = \tau^{2}\hat{s}_{n-1\; n}$.  Using this scaling
property we decompose the scattering equations based on their scaling
property in soft limit as,
\begin{eqnarray}\label{ds-scat-eqn-k2}
  E_a & = & \sum\limits_{\substack{b=1\\b \ne a}}^{n-2}\frac{\eso{a}{b}}
  {x_a-x_b} = 0, \qquad a\in \{1,2,\ldots n-2\}\nonumber\\
  E_{n-1} & = & \sum\limits_{b=1}^{n-2}\frac{\eso{n-1}{b}}{x_{n-1}-x_b}
                + \frac{\eso{n-1}{n}}{x_{n-1} -x_n} = 0, \nonumber\\
  E_n & = & \sum\limits_{b=1}^{n-2}\frac{\eso{n}{b}}{x_n-x_b}
            - \frac{\eso{n-1}{n}}{x_{n-1} - x_n} = 0.
\end{eqnarray}
The integral representation of the soft factor is,
\begin{equation}
  \mathtt{S}^{(2)}_{\text{DS}} = \int dx_n\; \delta(E_n) \int dx_{n-1} \;
  \delta(E_{n-1}) \left[\frac{x_{n-1}-x_1}{(x_{n-2}-x_{n-1})
      (x_{n-1}-x_n)(x_n-x_1)}\right]^2\ .
\end{equation}
The solutions to Eq.(\ref{ds-scat-eqn-k2}) fall into two
categories\cite{DoubleSoftPRD}, the non-degenerate solutions where
$|x_{n-1} - x_{n}| \sim \mathcal{O}\left(\tau^{0}\right)$ and the
degenerate solutions where
$|x_{n-1} - x_{n}| \sim \mathcal{O}\left(\tau\right)$.
In the adjacent soft limit, contribution from the degenerate
solutions dominate over those of the non-degenerate ones.  Since we are
interested in picking up the leading contribution, we will consider
only the degenerate solutions.

In the degenerate case we make a change of variables,
\begin{equation}
x_{n-1} = \rho + \xi, \qquad x_n = \rho - \xi,
\end{equation}
where, $\xi$ is $\mathcal{O}(\tau)$.  With this change of variables it
is convenient to re-express the delta functions with arguments
$E_{n-1}$ and $E_{n}$ in terms of sum and difference of the two
scattering equations. The integral representation of the double soft
factor then becomes,
\begin{eqnarray}
  \mathtt{S}^{(2)}_{\text{DS}}
  & = & \int d\rho \int d\xi \; \delta(E_{n-1}+E_n) \delta(E_{n-1}-E_n)
        \left[\frac{x_{n-2}-x_1}{\xi(x_{n-2}-\rho)
        (\rho-x_1)}\right]^2\ .
\end{eqnarray} 
The $\xi$ integral can be evaluated using the second delta function to
localise $\xi$ to its solution\cite{DoubleSoftPRD, VolovichZlotnikov,
  Saha:2016kjr, Saha:2017yqi, Chakrabarti:2017zmh}. This method is not
convenient when we study higher $k$ generalisations. Therefore instead
of solving for $\xi$, we can use the second delta function to convert
$\xi$ integration to a contour integral.  We can then use the contour
deformation method and pick up poles from the integrand which came out
of the Parke-Taylor factor in the soft limit.  The pole is seen to be
at $\xi = 0$ and in the neighbourhood of this pole we can approximate
$E_{n-1} - E_{n} \approx \frac{s_{n-1\; n}}{\xi}$. The soft factor
can now be evaluated as,
  \begin{eqnarray} \label{Adj-DS-2}    
    \mathtt{S}^{(2)}_{\text{DS}} & = & -\int d\rho \; \delta(E_{n-1} + E_n)
                                       \oint\limits_{\{\xi \to 0\}}
        \frac{d\xi}{\frac{\eso{n-1}{n}}{\xi}}\frac{1}{\xi^2}\left[
        \frac{x_{n-2}-x_1}{(x_{n-2}-\rho)(\rho-x_1)}
        \right]^2 \nonumber\\
  & = & \frac{1}{\eso{n-1}{n}}\oint\limits_{\{\rho\to x_{n-2}, x_1\}}
        \frac{d\rho}{\sum\limits_{b=1}^{n-2}\frac{\eso{n-1}{b} + \eso{n}{b}}
        {\rho-x_b}} \left[\frac{x_{n-2}-x_1}
        {(x_{n-2}-\rho)(\rho-x_1)}\right]^2 \nonumber\\
  & = & \frac{1}{\eso{n-1}{n}}\left[\frac{1}{\eso{n-1}{n-2} + \eso{n}{n-2}} +
        \frac{1}{\eso{n-1}{1} + \eso{n}{1}} \right].
\end{eqnarray}
It follows from the last line of \eq{Adj-DS-2} that
$\mathtt{S}^{(2)}_{\text{DS}}$ scales as $\tau^{-3}$. This scaling can
be also understood from the Feynman diagrams given below, where the
blob stands for the $n-3$ point tree diagram.
\begin{equation*}
\begin{aligned}
\begin{tikzpicture}[line width=1.5pt]
 \fill[black, thick](0,0) circle (0.6cm);
 \draw(-2.8,0) -- (-0.6, 0);
 \draw(-1.5,0) -- (-1.5, -1);
 \draw(-1.5, -1) -- (-2.3, -1.8);
 \draw(-1.5, -1) -- (-0.7, -1.8);
 \node at (-3,0.2) {$p_{n-2}$};
 \node at (-2.5, -2) {$k_{n-1}$}; 
 \node at (-.5, -2) {$k_{n}$};
 \end{tikzpicture} \\
 \textcolor{blue}{\frac{1}{s_{n-1\; n}}\times \frac{1}{s_{n-1\; n-2} +
     s_{n\; n-2}}}
 \end{aligned} 
 \qquad \qquad
 \begin{aligned}
 \begin{tikzpicture}[line width=1.5pt]
 \fill[black, thick](0,0) circle (0.6cm);
 \draw(0.6,0) -- (2.8, 0);
 \draw(1.5,0) -- (1.5, -1);
 \draw(1.5, -1) -- (2.3, -1.8);
 \draw(1.5, -1) -- (0.7, -1.8);
 \node at (3, 0.2) {$p_{1}$};
 \node at (2.5, -2) {$k_{n}$}; 
 \node at (0.5, -2) {$k_{n-1}$};
 \end{tikzpicture} \\
 \textcolor{blue}{\frac{1}{s_{n-1\; n}}\times \frac{1}{s_{n-1\; 1} + s_{n\; 1}}}
 \end{aligned} 
\end{equation*}
As can be seen from the above diagrams, the intermediate propagator
connecting two soft legs with momenta $k_{n-1}$ and $k_{n}$ with the
hard leg having the momentum $p_{a}$, $a=\{n-2,1\}$ gives a factor of
$\tau^{-2}$ whereas the propagator joining the blob gives a factor of
$\tau^{-1}$ making overall scaling of the double soft factor to be
$\tau^{-3}$.  This scaling may change in theories with the derivative
couplings, as factors of soft momenta can also come from the
interaction vertices. The power of soft momenta in the denominator is,
therefore, reduced and the double soft factors scale as $\tau^{-m}$
where $m\le 2$ \cite{DoubleSoftPRD, VolovichZlotnikov, Klose:2015xoa,
  Georgiou:2015jfa, Low:2015ogb, Saha:2016kjr, Saha:2017yqi,
  Chakrabarti:2017ltl, Chakrabarti:2017zmh, AtulBhatkar:2018kfi,
  Jain:2018fda, Marotta:2020oob}.

\subsubsection{Non-adjacent soft limit}
\label{sec:nasl}

We will now consider soft limits for two non-adjacent external states.
To illustrate this, we consider $n-2$ and $n$ to be soft punctures.
Following from our earlier analysis, it is easy to see that
$s_{n-2\; a}$ and $s_{n\; a}$ scale as $\tau$, and
$s_{n-2\; n}\to \tau^{2}$ in the limit $\tau\to 0$. The label
$a=1,\cdots, n-3, n-1$ denotes the hard external particles. The soft
factor can then be written as,
\begin{align}
  \label{eq:nonadj}
  \mathtt{S}^{(2)}_{\text{DS}}
  = & \int dx_n\; \delta(E_n) \int dx_{n-2}\; \delta(E_{n-2})
        \left[\frac{(x_{n-3}-x_{n-1})(x_{n-1}-x_1)}
        {(x_{n-3}-x_{n-2})(x_{n-2}-x_{n-1})
        (x_{n-1}-x_n)(x_n-x_1)}\right]^2 \nonumber\\
  = & \left(\frac{1}{\eso{n}{n-1}} + \frac{1}{\eso{n}{1}}\right)
        \left(\frac{1}{\eso{n-2}{n-3}} + \frac{1}{\eso{n-2}{n-1}}\right).
\end{align}
It readily follows that $\mathtt{S}^{(2)}_{\text{DS}}$ in
\eq{eq:nonadj} scales as $\tau^{-2}$ and is equal to the product of two
single soft factors found in Sec.(\ref{Sec:single-soft-2}). In general
if we consider soft limits in any two non-adjacent states, then the
double soft factor is the product of individual single soft factors
for the corresponding soft external states.


\section{Single Soft Theorem for Generalised Biadjoint Scalars}
\label{singlesoft3n}

In the previous section, we reviewed aspects of $k=2$ biadjoint scalar
theory in the CHY formalism
\cite{Cachazo:2013iaa,Cachazo:2013gna,Cachazo:2013hca,Cachazo:2013iea}.
We will now discuss its generalisation where instead of studying
punctures on $\cp{1}$ as will consider punctures on
$\cp{k-1}$\cite{Cachazo:2019ngv}.  From now on, we will refer to
external states as punctures on $\cp{k-1}$ since there is no clear
representation of them in terms of momenta.

The scattering equations for any $k$ are obtained by saddle point
analysis of the scattering potential function, which is a
generalisation of the $k=2$ scattering potential, given as,
\begin{align}
  \mathcal{S}^{(k)}=\sum_{a_1<a_2<\cdots <a_k}s_{{a_1a_2\cdots a_k}}
  \log(a_1a_2\cdots a_k),
\end{align}
where $(a_1a_2\cdots a_k)$ is the determinant,
\begin{align}
(a_1a_2\cdots a_k)=\begin{vmatrix}
\sigma_{a_1}^{(1)} & \sigma_{a_2}^{(1)} & \cdots &\sigma_{a_{k}}^{(1)}\\ 
\sigma_{a_1}^{(2)} & \sigma_{a_2}^{(2)} & \cdots &\sigma_{a_{k}}^{(2)}\\
\vdots & \vdots & \ddots &\vdots \\
\sigma_{a_1}^{(k)} & \sigma_{a_2}^{(k)} & \cdots &\sigma_{a_{k}}^{(k)}
\end{vmatrix},
\end{align}
and $\{\sigma_{a}^{(1)},\sigma_{a}^{(2)},\cdots, \sigma_{a}^{(k)}\}$
are the homogeneous coordinates describing the puncture $\sigma_a$ on
$\cp{k-1}$. The generalised Mandelstam variables, denoted as
$s_{a_1a_2\cdots a_k}$, are symmetric tensors of rank $k$ and has the
property that it vanishes whenever two indices on it are same,
\begin{align}
s_{a_1a_2\cdots a_i\cdots a_i\cdots a_k}=0,
\end{align}
and projective invariance of the scattering potential gives,
\begin{align}
  \sum_{a_2<a_3<\cdots <a_k \neq a_1}s_{a_1a_2\cdots a_k}=0\ ,\quad \forall a_1\ .
\end{align}
These two conditions are generalisations of masslessness condition and
momentum conservation of Mandelstam variables to higher $k$.

We will be working in the inhomogeneous coordinates, where the
scattering potential function is written as,
\begin{align}
  \mathcal{S}^{(k)}=\sum_{a_1<a_2<\cdots< a_k}s_{{a_1a_2\cdots a_k}}\log|a_1a_2
  \cdots a_k|,
\end{align}
where  $|a_1a_2\cdots a_k|$ is the determinant,
\begin{align}
|a_1a_2\cdots a_k|=\begin{vmatrix}
1 & 1 & \cdots &1\\ 
x_{a_1}^{1} & x_{a_2}^{1} & \cdots &x_{a_{k}}^{1}\\
\vdots & \vdots & \ddots &\vdots \\
x_{a_1}^{k-1} & x_{a_2}^{k-1} & \cdots &x_{a_{k}}^{k-1}\\
\end{vmatrix},
\end{align}
and $\{1,x_{a}^1,x_{a}^2,\cdots,x_{a}^k\}$ are the inhomogeneous
coordinates for the puncture $\sigma_a$. The scattering equation for
arbitrary $k$ can now be given as,
\begin{align}
  E_{a}^{(i)} & = \frac{\partial \mathcal{ S}^{(k)}}{\partial x_a^{i}}
                \nonumber \\
              &= \sum\limits_{a_2<\cdots< a_k \ne a} \frac{s_{aa_2\cdots a_k}}
                {|aa_2\cdots a_k|}\frac{\partial}{\partial x_{a}^{i}}
                |aa_2\cdots a_k| = 0, \qquad \forall a=1,\cdots,n.
\end{align}
This equation hold for each $i=1,\cdots k-1$.  Analogous to the $k=2$
case, the biadjoint scalar amplitude for general $k$ is constructed as
\cite{Cachazo:2019ngv},
\begin{equation}
  \label{eq:nptamp}
  m_n^{(k)}(\alpha|\beta) = \int d\mu_n \text{PT}^{(k)}[\alpha]
  \text{PT}^{(k)}[\beta]\ ,
\end{equation}
where $\alpha$ and $\beta$ are specific ordering of $n$ scalars in
this amplitude, and PT stands for the Parke-Taylor factor, which for
fixed $k$ takes the form,
\begin{equation}
  \label{eq:ptk3}
  \mathrm{PT}^{(k)}[1,2,\cdots,n]:=\frac{1}{|1\;2\;\cdots\;k||2\;3\;
    \cdots\;k-1|\cdots|n\;1\;\cdots\;k-1|}\ .
\end{equation}
Two copies of the PT factors signifies that the scalar is in the
biadjoint representation.  We will work with $\alpha=\beta=\mathtt{I}$.

In this section, we will study soft theorems for the generalised
biadjoint scalars for any $k$.  We will begin by recalling the single
soft theorem, which was studied in \cite{Sepulveda:2019vrz}. We will
first review the case for $k=3$ theory and then summarise the general
$k$ case briefly. Through this review, we will set up the notation,
which will be useful when we study the double soft theorem.

\subsection{Single soft theorems for $k=3$ amplitudes}

Let us consider the $n$-point amplitude with $n$-th state going soft.
A convenient way to deal with the soft limit is to express the
$n$-point function with one soft state in terms of $(n-1)$-point
function.  This is done by extracting the terms in the PT factor which
depend on the puncture corresponding to the $n$-th particle.  After
extracting those terms, the remaining PT factor is almost that for the
$n-1$ point function.  However, to ensure the cyclic symmetry of the
PT factor we need to multiply and divide by terms which reinstate the
cyclic symmetry of the $(n-1)$-point PT factor. Taking this into account
we can write,
\begin{equation}
\label{PTn2n-1}
\text{PT}^{(3)}[1,2,\cdots, n] = \frac{\td{n-2}{n-1}{1}
	\td{n-1}{1}{2}}{\td{n-2}{n-1}{n}\td{n-1}{n}{1}\td{n}{1}{2}}
\text{PT}^{(3)}[1,2,\cdots\,n-1].
\end{equation}
On the RHS of \eq{PTn2n-1}, the PT factor involves $n-1$ punctures and
the factor multiplying it has two parts, while the denominator
contains the soft punctures extracted from the $n$-point PT factor,
the numerator terms are remnants of the cyclic symmetry of the
$n-1$-point function. In order to take the soft limit, we introduce a
parameter $\tau$ and introduce the limit,
\begin{equation}
\label{gmsoft}
\ese{n}{a}{b} = \tau \hat{s}_{n\,a\,b} \rightarrow 0, \qquad \forall a,b \ne n\ ,
\end{equation}
where $\ese{a}{b}{c}$ is a symmetric third rank tensor whose
components are the generalised Mandelstam variables for $k=3$
case. The limit taken above is akin to the limit
$s_{a\;b}\rightarrow \tau \hat{s}_{a\;b}$ in the $k=2$ case where
$s_{a\;b}$ are the Mandelstam variables for $k=2$ which can be
expressed in terms of the kinematic variables.

In this limit scattering equations decouple at the leading order,
since $\ese{a}{b}{n}$ scales differently compared to $\ese{a}{b}{c}$
when $\{a,b,c\}\not=n$.  The decoupled equations take the
form\footnote{\label{foot1}In fact, in the $E_a^{(i)}$ equation there
  is an additional term
  $\tau\frac{\hat{s}_{a\,b\,n}}{\td{a}{b}{n}} \frac{\partial}{\partial
    x_a^i}\td{a}{b}{n}$. When one considers the singular configuration
  $\td{a}{b}{n} \sim \tau$, which can happen when $a,b,n$ are
  collinear or $n$-th puncture collides with either puncture $a$ or
  puncture $b$, then the derivative term is of the form
  $\begin{vmatrix} 1 && 1\\ x_{b}^j && x_{n}^j
  \end{vmatrix}$, where $j \ne i$. This determinant, in general, may
  not scale to 0 as $\tau\to 0$.  Hence in the case of the singular
  solution this additional term is of $\mathcal{O}(1)$, nevertheless
  its contribution appears at subleading order in the soft
  theorem\cite{Cachazo:2019ble}.  We will ignore these configurations
  as we are interested in the leading order results, which have
  contributions only from the regular solutions.},
\begin{align}
	\label{eq:gsceq}
	E_n^{(i)}
	= &\ \tau\sum\limits_{1\le a < b \le n-1} \frac{\hat{s}_{a\,b\,n}}
	{\td{a}{b}{n}}\frac{\partial}{\partial x_n^i}
	\td{a}{b}{n} = 0\ ,\nonumber \\
	& \\
	E_a^{(i)}
	= & \sum\limits_{\{b,c\} \ne\{a,n\}} \frac{\ese{a}{b}{c}}{\td{a}{b}{c}}
	\frac{\partial}{\partial x_a^i}\td{a}{b}{c}
	= 0, \qquad \forall a \ne n, \quad i= 1,2\ .\nonumber
\end{align}
In the soft limit we replace the $n$-point PT factor by the
$n-1$-point PT factor multiplied by the soft factor extracted from the
original PT factor as given in \eq{PTn2n-1}.  With this substitution
and explicitly pulling out the integration over the soft kinematic
variables, \eq{eq:nptamp} takes the form,
\begin{align}
	\label{eq:nptk3}
	m_n^{(3)} (\mathtt{I}| \mathtt{I}) =
	& \int d\mu_{n-1} \prod\limits_{i=1}^2\int dx_n^i \delta(E_n^{(i)})
	\left(\frac{\td{n-2}{n-1}{1}\td{n-1}{1}{2}}{\td{n-2}{n-1}{n}
		\td{n-1}{n}{1}\td{n}{1}{2}}\right)^2\nonumber\\
	\times & \left(\text{PT}^{(3)}[1,2,\cdots, n-1]\right)^2\\[3mm]
  := &\ \mathtt{S}^{(3)}_n m_{n-1}^{(3)}(\mathtt{I}| \mathtt{I})
       \nonumber\ ,
\end{align}
where,
\begin{equation}
\label{eq:softf}
\mathtt{S}^{(3)}_n = \prod\limits_{i=1}^2\int dx_n^i \delta(E_n^{(i)})
\left(\frac{\td{n-2}{n-1}{1}\td{n-1}{1}{2}}{\td{n-2}{n-1}{n}
	\td{n-1}{n}{1}\td{n}{1}{2}}\right)^2\ .
\end{equation}
The integration over $x_n^i$ in \eq{eq:softf} implements the
scattering equations \eq{eq:gsceq} through the $\delta$-functions.
Equivalently, it is convenient to replace the $\delta(E_{n}^{(i)})$ by
poles located at the zeros of the scattering equations.  This is
achieved by removing the $\delta$-functions and putting the scattering
equations in the denominator of the integrand.  This allows us to use
the contour integral method, and we can compute the integral by
deforming the contour away from the zeros of the scattering equation.
The residues collected from the poles away from the zeroes of the
scattering equation give identical contribution except with the
opposite sign.  Since we will be working with multiple complex
variables, two per puncture in the projective coordinates, as in
\eq{eq:nptk3} or three per puncture in the homogeneous coordinates, it
is suitable to use the global residue theorem \cite{Griffiths:433962}
to carry out the contour deformation.

Since the contribution from the scattering equations is not picked up
in the contour deformation, only possible contributions come from the
soft factors in \eq{eq:softf}.  These contributions can be categorized
into two types, collision singularities and collinear singularities.
The collision singularities are those where the puncture $\sigma_n$
corresponding to the soft state collides with one of the hard
punctures in the denominator of the integrand of \eq{eq:softf}.  The
collinear singularities on the other hand, correspond to two punctures
becoming collinear with the soft punctures in the denominator of the
integrand of \eq{eq:softf}.  We will first discuss the contribution
from collision singularities and then discuss the contribution from
collinear singularities.

\subsubsection{Collision singularities}
\label{sec:clsnsngl}

The collision singularities in \eq{eq:softf} occur for the following
two planar configurations: (i) $\sigma_n\to \sigma_1$, (ii)
$\sigma_n\to\sigma_{n-1}$.  In the first case we parametrize the
$n$-th variable as\footnote{For the $k=3$ case, we will use
  $\{1,x_a^1,x_a^2\}$ or $\{1,x_a,y_a\}$ interchangeably, to denote the
  inhomogeneous coordinates of the $\sigma_a$ puncture.},
\begin{equation}
x_n = x_1 + \epsilon, \qquad y_n = y_1 + \epsilon\alpha, \quad
\epsilon \to 0\ ,
\end{equation}
where we have chosen to work in the inhomogeneous coordinates.  With
this parametrisation the integration measure over the soft puncture
variables becomes,
\begin{equation}
\prod\limits_{i=1}^{2}dx_n^{i}= dx_{n}dy_{n} = \epsilon
d\epsilon d\alpha,
\end{equation}
and delta functions, treated as the top form on $\sigma_n$, transforms
in the following way,
\begin{equation}
  \delta^{(2)}\left(\partial_{x_{n}}\mathcal{ S}^{(3)}, \partial_{y_{n}}
    \mathcal{ S}^{(3)}\right) = \epsilon \delta^{(2)}\left(\partial_{\epsilon}
    \mathcal{ S}^{(3)}, \partial_{\alpha}\mathcal{ S}^{(3)}\right).
\end{equation}
In deriving this relation we have used the fact that the change of
variables implemented for the argument of the $\delta$-function
generates a term proportional to $\epsilon^{-1}$, and
$\delta(ax)= |a|^{-1}\delta(x)$.
The scattering equations become,
\begin{align}
  d \mathcal{ S}^{(3)}
  &= \tau\sum\limits_{2\le b\le n-1}\hat{s}_{1\;b\;n}
    \left(\frac{d\epsilon}{\epsilon} +
    \frac{d\alpha}{\alpha - \alpha_{b}}\right)
    \nonumber\\
  \Rightarrow \quad \partial_{\epsilon}\mathcal{ S}^{(3)}
  & = \frac{\tau}{\epsilon}\sum_{a=2}^{n-1}
    \hat{s}_{1\;a\;n}, \qquad \partial_{\alpha}\mathcal{ S}^{(3)} = \tau
    \sum\limits_{2\le b\le n-1}\frac{\hat{s}_{1\;b\;n}}{\alpha - \alpha_{b}}\ ,
\end{align}
where, $\alpha_b$ is the $\cp{1}$ projection of $\sigma_b$.
The soft factor can then be evaluated as,
\begin{align}
  \mathtt{S}^{(3)\perp}_{n}(n;1)
  &= \frac{1}{\tau^{2}}\oint \frac{\epsilon^{2} d\epsilon}{\frac{1}{\epsilon}
    \sum\limits_{a=2}^{n-1}\hat{s}_{1\;a\;n}} \oint \frac{d\alpha}
    {\sum\limits_{b=2}^{n-1}\frac{\hat{s}_{1\;b\;n}}{\alpha-\alpha_{b}} }
    \left[\frac{\alpha_{2} - \alpha_{n-1}}{\underline{\epsilon^{2}
    \left(\alpha- \alpha_{n-1}\right)\left(\alpha - \alpha_{2}\right)}}
    \right]^{2} \nonumber\\
  &= \frac{1}{\sum\limits_{a=2}^{n-1}s_{1\;a\;n}} \left(\frac{1}{s_{1\;2\;n}}
    + \frac{1}{s_{n-1\; n\; 1}}\right),
\end{align}
where the integration in $\alpha$ variable is that corresponding to a
single soft factor for $k=2$ from \eq{chy-single-soft}.
A similar analysis in the case (ii), when
$\sigma_{n} \rightarrow \sigma_{n-1}$, leads the corresponding residue
to be equal to,
\begin{align}
  \mathtt{S}^{(3)\perp}_{n}(n;n-1)
  &=\frac{1}{\sum\limits_{a=1}^{n-2} s_{n-1\;n\;a}}
    \left(\frac{1}{s_{n-1\; n\; 1}} + \frac{1}{s_{n-2\; n-1\; n}}\right)\ .
\end{align}
Besides the planar collisions, we could also have non-planar
collisions, e.g., $\sigma_n\to \sigma_2$, $\sigma_n\to\sigma_{n-2}$.
Most of the analysis above carries through in these cases as well
except that in these cases we have only one determinant in the
denominator of the soft factor becoming proportional to $\epsilon$.
This gives us $\epsilon^2$ factor in the denominator, which is not
good enough to offset $\epsilon^3$ factor in the numerator and as a
result this type of degeneration does not contribute to the single
soft theorem.


\subsubsection{Collinear singularities}
\label{sec:colsing}

We will now look at the collinear singularities.  As mentioned
earlier, collinear singularities occur when the soft particle $n$
becomes collinear with two hard punctures.  We will begin with the
case where the soft puncture of $n$ becomes collinear with hard
punctures of $n-2$ and $n-1$.  In the homogeneous coordinates, we
have,
\begin{align}
	\sigma_n=\alpha\sigma_{n-1}+\xi,
\end{align}
where $\xi$ lies on the straight line that connects the punctures
$n-1$ and $n-2$. In this case, the determinant $\tp{n-2}{n-1}{n}$
becomes,
\begin{align}
  \label{eq:ptpoles}
	\tp{n-2}{n-1}{n}&=\tp{n-2}{n-1}{\xi}=0,
\end{align}
where we have used the fact that $\xi$ is collinear with $n-2$ and
$n-1$. The \eq{eq:ptpoles} corresponds to poles of the PT
factor. However note that a soft puncture becoming collinear with two
hard punctures is a codimension one singularity.  Whereas for the
$\cp{2}$ integration to give non-zero residue, one needs a codimension
two singularity. Such a codimension two singularity occurs when the
soft puncture becomes simultaneously collinear with two sets of hard
punctures.  We will therefore consider the case where soft puncture
$n$ becomes collinear with two pairs of hard punctures $n-2,n-1$ and
$1,2$.  This can be parametrised as,
\begin{align}
	\sigma_n=\alpha\sigma_{n-1}+\beta\sigma_1+\xi,
\end{align}
where, as $\beta\to 0$ we obtain the straight line corresponding to the
punctures $n-2,n-1$ and, as $\alpha\to 0$ we obtain the straight line
corresponding to the punctures $1,2$. The above equation when written
in inhomogeneous coordinates reads,
\begin{align}\label{collinear-inhomogeneous}
	x_n&=\alpha x_{n-1}+\beta x_1+(1-(\alpha+\beta))x_\xi\nonumber\\
	y_n&=\alpha y_{n-1}+\beta y_1+(1-(\alpha+\beta))y_\xi,
\end{align}
which makes the determinants in the PT factor transform as,
\begin{align}
	\td{n-2}{n-1}{n}&=\beta\td{n-2}{n-1}{1},\nonumber\\
	\td{n}{1}{2}&=\alpha\td{n-1}{1}{2},\nonumber\\
	\td{n-1}{n}{1}&=\td{n-1}{\xi}{1}.
\end{align}
The measure, on the other hand, transforms as,
\begin{equation}
dx_{n}dy_{n} = \td{n-1}{1}{\xi} d\alpha d\beta\ ,
\end{equation}
and the scattering equations become,
\begin{eqnarray}
  d\mathcal{ S}^{(3)}
  & = & \frac{s_{n-2\; n-1\; n}}{\alpha}d\alpha +
        \frac{s_{n\;1\;2}}{\beta}d\beta \nonumber\\
  \Rightarrow \partial_{\alpha}\mathcal{ S}^{(3)}
  & = & \frac{s_{n-2\; n-1\; n}}{\alpha}, \qquad \partial_{\beta}
        \mathcal{ S}^{(3)} = \frac{s_{n\;1\;2}}{\beta}.
\end{eqnarray}
The transformation rule \eq{collinear-inhomogeneous} implies, 
\begin{equation}
  \delta^{2}\left(\partial_{x_{n}}\mathcal{ S}^{(3)}, \partial_{y_{n}}
    \mathcal{ S}^{(3)}\right)  =  \td{n-1}{1}{\xi} \delta^{2}
  \left(\partial_{\alpha}\mathcal{ S}^{(3)}, \partial_{\beta}
    \mathcal{ S}^{(3)}\right).
\end{equation}
The soft factor then takes the form, 
\begin{eqnarray}
  \mathtt{S}_n^{(3)\parallel}
  & = & \int\frac{(\td{n-1}{1}{\xi})^{2}d\alpha d\beta}
        {\frac{s_{n-2\; n-1\; n}}{\alpha}\times\frac{s_{n12}}{\beta}}
        \left(\frac{1}{\underline{\alpha\beta} |n-1\; \xi\; 1|}\right)^{2}
        \nonumber\\
& = & \frac{1}{s_{n-2\; n-1\; n}\: s_{n\;1\;2}}.
\end{eqnarray}
Therefore the full single soft factor for $k=3$ biadjoint scalar
theory is given by,
\begin{align}
  \mathtt{S}_n^{(3)}
  &=\mathtt{S}_n^{(3)\perp}(n;1)+
    \mathtt{S}_n^{(3)\perp}(n-1;n)+\mathtt{S}_n^{(3)\parallel}
    \nonumber\\
  &=\frac{1}{\sum\limits_{a=2}^{n-1}s_{1\;a\;n}}
    \left(\frac{1}{s_{1\;2\;n}} + \frac{1}{s_{n-1\; n\; 1}}
    \right)+\frac{1}{\sum\limits_{a=1}^{n-2}s_{n-1\;n\;a}}
    \left(\frac{1}{s_{n-1\; n\; 1}} + \frac{1}{s_{n-2\; n-1\; n}}
    \right)\nonumber\\
  & \hspace{5cm}\quad+\frac{1}{s_{n-2\; n-1\; n}\: s_{n12}}.
\end{align}
Thus the single soft factor for $k=3$ is obtained by studying the
boundary structure in the moduli space which comprises of both
collision and collinear type singularities of codimension two.

\subsection{Single soft limit for arbitrary $k$}
\label{sec: gen-k-single-soft}

The above discussions for single soft theorem can be generalised for
arbitrary $k$. We consider soft limit in $n$-th external state, such
that $s_{a_{1}a_{2}\cdots a_{k-1}n}$ scales as $\tau$ with
$\tau\rightarrow 0$ for any $a_{i} \in \{1,2, \cdots n-1\}$. The
scattering equations can then be decomposed in the following way:
\begin{eqnarray}
  E_{a_{1}}^{(i)}
  & = & \sum\limits_{\substack{1\le a_{2} < a_{3} < \cdots < a_{k}\le n-1
  \\ a_{2},\cdots, a_k\ne a_{1}}} \frac{s_{a_{1}\,a_{2}\,\cdots\, a_{k}}}
  {|a_{1}\,a_{2}\,\cdots\, a_{k}|} \frac{\partial}{\partial x_{a_{1}}^i}
  |a_{1}\,a_{2}\,\cdots\, a_{k}|=0, \qquad \forall a_{1} \nonumber\\
  E_{n}^{(i)}
  & = & \sum\limits_{1\le a_{1}< \cdots <a_{k-1}\le n-1}
        \frac{s_{a_{1}\,\cdots\, a_{k-1}\,n}}{|a_{1}\,\cdots\, a_{k-1}\,\sigma_{n}|}
        \frac{\partial}{\partial x_{n}^{i}}|a_{1}\,\cdots\, a_{k-1}\,\sigma_{n}|
        =0,
\end{eqnarray}
where $i = 1,2, \cdots k-1$. Here we consider only the regular
solutions to scattering equations and hence we neglect any
$\mathcal{O}\left(1\right)$ terms that may arise when
$|a_{1}\cdots a_{k-1}\sigma_{n}|\sim \tau$.

The soft factor for arbitrary $k$ can be expressed in the integral form as, 
\begin{eqnarray}\label{general-k-soft}
  \mathtt{S}^{(k)}
  & = & \int \prod\limits_{i=1}^{k-1}dx_{n}^{i}\; \delta^{(k-1)}
        \left(E_{n}^{i}\right)\nonumber\\
  && \times \left[\frac{|(n-k+1)\cdots 1||(n-k+2)\cdots 1\, 2|\cdots
     |n-1\,1\,2\cdots\, k-1|}{|(n-k+1)\,\cdots\, n||(n-k+2)\cdots n\,1|
     \cdots |n\,1\,2\cdots\,k-1|}\right]^{2}\ .\nonumber\\
\end{eqnarray}
In \cite{Sepulveda:2019vrz}, the expression in \eq{general-k-soft} was
evaluated in terms of the generalised Mandelstam variables by an
iterative procedure and a prescription for calculating the soft factor
for any given $k$ was presented. The scaling of the soft factor in
this case can be seen to be $\tau^{-(k-1)}$. For the purposes of this
work, the formal factorisation given in \eq{general-k-soft} is
sufficient and we refer the reader to \cite{Sepulveda:2019vrz} for
further details on the single soft factor.


\section{Double Soft Theorem for $k=3$ Amplitudes}
\label{sec:dstg3n}

We will now look at the double soft theorems, where we have two
external states becoming soft.  Here we have two main cases, and we
will treat them one at a time.  To begin with, we will discuss the
simultaneous double soft limit.  In this case, both the external states
are going soft at the same rate, and therefore the limits cannot be taken
independently.  The simultaneous double soft theorem has multiple
sub-cases, and we will analyse each sub-case separately.  The other kind
of double soft limit is called the consecutive double soft limit,
where one state goes soft at a faster rate than the other soft state.
This limit clearly has a hierarchical structure and is relatively easy
to deal with, and we will take up this case later.


\subsection{Simultaneous double soft limit}
\label{sec:sdsl}

Let us now look at the simultaneous double soft limit.  This limit
implies two external states are going soft at the same rate,
parametrised in terms of $\tau$, which in the soft limit will be taken
to zero.  In this case, we can have several configurations which
qualify as double soft limit, but their treatment differs.  For
example, we can have adjacent external states going soft, which is the
case we will deal with first.  However, besides this, we can have next
to adjacent states becoming soft.  The singularity structure of this
case is quite different from the case of adjacent external states
going soft.  We could also have the next to next to adjacent states in
the scattering going soft.  For the case at hand, namely $k=3$, this
choice of configurations with soft scattering states further away from
each other, resulting in an expression which is a product of two
single soft factors.  In fact, for arbitrary $k$, if there are at
least $k-1$ hard states between two soft states, then the double soft
limit is just a product of two single soft factors.

\paragraph{Soft Limits for Adjacent Particles:}
\label{sec:APS}
We will consider $n$-th and $(n-1)$-th external states
going soft.   This limit corresponds to the
following behaviour of the generalised Mandelstam variables,
\begin{equation}
  \ese{a}{b}{n-1} \to \tau \hat{s}_{a\,b\,n-1},\quad \ese{a}{b}{n}\to
  \tau\hat{s}_{a\,b\,n},\quad \ese{a}{n-1}{n}\to\tau^{2}\hat{s}_{a\,n-1\,n},
  \quad a,b =1,2,\cdots, n-2.
\end{equation}
Here we will be concerned with the leading order soft
factorisation. Therefore it suffices to consider only regular
solutions to the scattering equations, \textit{i.e.}, we assume none
of the determinants $\td{a}{b}{n}$ and $ \td{a}{b}{n-1}$ scale as
$\mathcal{O}\left(\tau\right)$
$ \forall a,b \in \{1,2, \cdots , n-2\}$. In \cite{Cachazo:2019ble} it
has been shown that singular solutions to scattering equations for
$k\ge 3$ contribute to the subleading soft theorem.  Therefore these
solutions will not be part of our analysis.  The scattering equations
to the leading order can be written as,
\begin{align}
  \label{sceq2sft}
  E_{a}^{(i)}
  & = \sum\limits_{b,c \ne a, n-1, n}\frac{\ese{a}{b}{c}}{\td{a}{b}{c}}
    \frac{\partial}{\partial x_{a}^{(i)}}\td{a}{b}{c} = 0\ , \qquad
    \forall a \nonumber\\
  E_{n-1}^{(i)}
  & = \tau \sum\limits_{a,b\ne n-1,n}\frac{\hat{s}_{a\,b\,n-1}}{\td{a}{b}{n-1}}
    \frac{\partial}{\partial x_{n-1}^{(i)}}\td{a}{b}{n-1} + \tau^{2}
    \sum\limits_{a=1}^{n-2} \frac{\hat{s}_{a\,n-1\,n}}{\td{a}{n-1}{n}}
    \frac{\partial}{\partial x_{n-1}^{(i)}}\td{a}{n-1}{n} =0\ , \nonumber\\
  E_{n}^{(i)}
  & = \tau \sum\limits_{a,b\ne n-1,n}\frac{\hat{s}_{a\,b\,n}}{\td{a}{b}{n}}
    \frac{\partial}{\partial x_{n}^{(i)}}\td{a}{b}{n} +
    \tau^{2}\sum\limits_{a=1}^{n-2}\frac{\hat{s}_{a\, n-1\,n}}{\td{a}{n-1}{n}}
    \frac{\partial}{\partial x_{n}^{(i)}}\td{a}{n-1}{n} = 0\ ,
\end{align}
where $i=1,2$ labels components of the inhomogeneous coordinates
of the puncture.

Let us first look at the integrand in \eq{eq:nptamp} with $(n-1)$-th and
$n$-th particle soft.  In the double soft limit, as was done in the
single soft limit, we extract the dependence on the soft punctures and
write the remaining factor as the Parke-Taylor factor for $n-2$
punctures.  The resulting expression can be written as,
\begin{equation}
  \label{PT2soft}
  \text{PT}^{(3)}[12\cdots n] = \frac{\td{n-3}{n-2}{1} \td{n-2}{1}{2}}
  {\td{n-3}{n-2}{n-1} \td{n-2}{n-1}{n} \td{n-1}{n}{1}\td{n}{1}{2}}
  \text{PT}^{(3)}[12\cdots n-2]\ .
\end{equation}
In the double soft limit, the amplitude can be written as,
\begin{align}\label{softk3}
  m_{n}^{(3)}(\mathtt{I}|\mathtt{I})
  &= m_{n-2}^{(3)}(\mathtt{I}|\mathtt{I})\int \prod\limits_{i}dx_{n-1}^{(i)}
    dx_{n}^{(i)}\delta(E_{n-1}^{(i)})\delta(E_{n}^{(i)}) \nonumber\\
  &\hspace{2.5cm}\times \left(\frac{\td{n-3}{n-2}{1} \td{n-2}{1}{2}}
    {\td{n-3}{n-2}{n-1}\td{n-2}{n-1}{n}\td{n-1}{n}{1}\td{n}{1}{2}}
    \right)^{2}\ .
\end{align}
The soft factor can be computed by deforming contours of integration
away from the original poles coming from scattering equations of soft
particles which are written in terms of the delta-functions in the
integral. In this process, we encounter poles in the integrand, which
occur when the determinants in the denominator of the integrand
vanish.  As in the single soft limit, the determinant vanishes in two
possible ways.  Either there is a collision singularity, that is when
two punctures collide or when three punctures become collinear.  In
the case of the double soft limit, we encounter more intricate
combinations of these two types of singularities.

The second and third equations in \eq{sceq2sft} have two sums. First
sum always scales linearly in $\tau$, however, depending on the
behaviour of $\td{a}{n-1}{n}$, the terms in the second sum can either
be linear or quadratic in $\tau$.  This gives rise to two types of
solutions to the scattering equations:
\begin{itemize}
\item Non-degenerate solutions: They correspond to
  $\td{a}{n-1}{n} \sim \mathcal{O}\left(\tau^{0}\right)$. In this
  case, the second sum of last two equations in \eq{sceq2sft} are
  sub-dominant compared to the first sum.
\item Degenerate solutions: They correspond to
  $\td{a}{n-1}{n} \sim \mathcal{O}\left(\tau\right)$. In this case,
  some or all the terms in the second sum are of $\mathcal{O}(\tau)$
  and hence are of the same order as the first sum.
\end{itemize}
For non-degenerate solutions the double soft factor scales as
$\tau^{-4}$. This behaviour can be understood from Eq.(\ref{softk3}) -
there are four delta functions containing scattering equations of soft
external states in the arguments, each of which contributes to a
factor of $\tau$ in the denominator, and the integrand is independent
of $\tau$. For degenerate solutions, the determinants which depend on
both $n-1$ and $n$ punctures contribute to a factor of $\tau$ each,
and there are two such determinants in the denominator of the
integrand. So the integrand scales as $\tau^{-4}$, and it can be
checked that measure goes as $\tau^{-2}$ making the overall scaling of
the double soft factor as $\tau^{-6}$. As we are interested in the
leading soft theorem ,we will only present the analysis of the
degenerate solutions in the following subsection.

\subsection{Degenerate solutions}

As argued above, similar to $k=2$ case studied in
sec.(\ref{Sec:double-soft-2}), degenerate solutions dominate in the
double soft limit for the adjacent particles going soft.  We will
therefore concentrate on this sector and analyse the leading singular
behaviour in the double soft limit.  In order to do that, let us first
look at the scattering equations in Eq.(\ref{sceq2sft}). For the
degenerate solutions denominator in the second sum of $E_{n-1}$ and
$E_{n}$ equations scale as $\tau$, making the overall scaling of the
terms to be $\tau$. There are two possible ways where the determinant
$|a\; n-1\; n| \sim \tau$:
\begin{itemize}
\item if the two soft punctures collide with each other,
  $\sigma_{n} - \sigma_{n-1} \sim\tau$.
\item if the two soft punctures are nearly collinear to any one of the
  hard punctures. This can happen if the straight line joining
  $\sigma_{n-1}$ and $\sigma_{n}$ is away from a hard puncture,
  $\sigma_{a}$ by a distance of
  $\mathcal{O}\left(\tau\right)$\footnote{There can not be more than
    two such hard punctures because in that case the two hard
    punctures, say, $a$ and $b$ will be collinear with a soft puncture
    in the limit $\tau \rightarrow 0$ making both
    $|a\; n-1\; n| \rightarrow 0$ and $|b\; n-1\; n| \rightarrow
    0$. But in the latter case, they are part of singular solutions and
    hence we do not consider them in this discussion.}.
\end{itemize}
We will analyse these two cases in detail in the following subsections.

\subsubsection{Collision of soft punctures}

We will begin with the analysis of the collision singularity.  It is
convenient to make the following change of variables in order to do
the integration:
\begin{equation}
x_{n-1}^i = \rho^i + \xi^i, \qquad x_n^i = \rho^i - \xi^i, \qquad i=1,2.
\end{equation}
where $\rho$ is $\mathcal{O}\left(1\right)$ and $\xi$ is
$\mathcal{O}(\tau)$. In component form they can be given by
$\rho = \begin{pmatrix} 1\\ \rho^{1} \\ \rho^{2}
\end{pmatrix}$ and $\xi = \begin{pmatrix}
0\\ \xi^1 \\ \xi^2
\end{pmatrix}$. In terms of these variables we can re-express the
scattering equations as,
\begin{align}\label{deg eq}
  E_{n-1}^{(i)} + E_n^{(i)}
  = & \sum\limits_{1\leq a<b \leq n-2} \frac{\ese{a}{b}{n-1} + \ese{a}{b}{n}}
      {\td{a}{b}{\rho}}\frac{\partial}{\partial\rho^i}\td{a}{b}{\rho}\ ,
      \nonumber\\
  E_{n-1}^{(i)} - E_n^{(i)}
  = & \sum\limits_{1\leq a<b\leq n-2} \frac{\ese{a}{b}{n-1} - \ese{a}{b}{n}}
      {\td{a}{b}{\rho}}\frac{\partial}{\partial\rho^i}\td{a}{b}{\rho} +
      \sum\limits_{a=1}^{n-2}
      \frac{\ese{a}{n-1}{n}}{\td{a}{\rho}{\xi}}\frac{\partial}{\partial\xi^i}
      \td{a}{\rho}{\xi}\ .
\end{align}
Determinant containing $a, \rho$ and $\xi$ then takes the form,
\begin{align}
|a\,\rho\,\xi | = & \begin{vmatrix}
1 & 1 & 0\\
x_a^1 & \rho^1 & \xi^1 \\
x_a^2 & \rho^2 & \xi^2
\end{vmatrix} \nonumber\\
  = & \xi^1(\rho^1-x_a^1)(\alpha - \alpha_{a}), \qquad
      \text{where we define} \quad \alpha= \frac{\xi^2}{\xi^1}\ ,
      \quad \alpha_{a} = \frac{x_a^2 - \rho^2}{x_a^1 - \rho^1}\ .
\end{align}
The measure transforms as,
\begin{equation}
  \prod\limits_{i=1}^2dx_{n-1}^idx_n^i\delta(E_{n-1}^{(i)})
  \delta(E_n^{(i)}) = 16\ d^2\rho\; d^2\xi \; \delta^{(2)}(E_{n-1}+ E_n)
  \delta^{(2)}(E_{n-1}-E_n)\ .
\end{equation}
The second delta function in the measure can be used to solve for
$\xi$ in terms of $\rho$ to localize the $\xi$ integration analogous
to the $k=2$ case\cite{DoubleSoftPRD, VolovichZlotnikov, Saha:2016kjr,
  Saha:2017yqi, Chakrabarti:2017zmh},
\begin{equation}
  \delta^{(2)}(E_{n-1}-E_n) = \sum\limits_{\xi_{0}}\frac{\delta^{(2)}
    (\xi-\xi_0)}
  {\begin{vmatrix}
      \frac{\partial(E^{(1)}_{n-1}-E^{(1)}_n)}{\partial\xi^1} &
      \frac{\partial(E^{(1)}_{n-1} - E^{(1)}_n)}{\partial\xi^2} \\
      \frac{\partial(E^{(2)}_{n-1}-E^{(2)}_n)}{\partial\xi^1} &
      \frac{\partial(E^{(2)}_{n-1}-E^{(2)}_n)}{\partial\xi^2}
\end{vmatrix}}\ ,
\end{equation}
where $\xi_{0}$ are the solutions of the scattering equations for
$\xi$.

However, we find that solving for $\xi$ from the scattering equations
is rather complicated as it leads to a high degree polynomial
equation.  Therefore we take an alternate approach analogous to the
one we adopted while discussing $k=2$ soft limit in
Sec.(\ref{Sec:double-soft-2}).

Instead of using the second delta function to localize $\xi$ integral,
we will use it to convert the $\xi$ integration to a contour
integral. We can express the soft factor as,
\begin{equation}\label{deg-integral}
  \mathtt{S}^{(3)}_{\text{deg}} = \int d^2\rho\; d^2\xi \; \delta^{(2)}
  (E_{n-1}+E_n) \delta^{(2)}(E_{n-1}-E_n) \left[\frac{\td{n-3}{n-2}{1}
      \td{n-2}{1}{2}}{\td{n-3}{n-2}{\rho}\td{n-2}{\xi}{\rho}
      \td{\rho}{\xi}{1}\td{\rho}{1}{2}}\right]^2.
\end{equation}
In $\mathtt{S}^{(3)}_{\text{deg}}$, the subscript implies that the
soft factor is related to the degenerate solutions. We note that,
\begin{equation}
  \td{n-2}{\xi}{\rho}\td{\rho}{\xi}{1} = (\xi^1)^2(\alpha-\alpha_{n-2})
  (\alpha-\alpha_{1})(x_{n-2}^1-\rho^1)(x_1^1-\rho^1)\ .
\end{equation}
Vanishing of the L.H.S. implies either $\xi^{1} \rightarrow 0$, and/or
$\alpha \rightarrow \alpha_{n-1}$, and/or $\alpha \rightarrow
\alpha_{1}$.  We consider the following change of variables,
\begin{eqnarray}\label{collision-limit}
 \xi^1 &= & \epsilon \nonumber\\
\Rightarrow  \xi^2 & = & \epsilon\alpha\ ,
\end{eqnarray}
which simplifies the integration measure,
\begin{equation}
d^2\xi = \begin{vmatrix}
1 & 0 \\
\alpha  & \epsilon
\end{vmatrix}
d\epsilon \; d\alpha = \epsilon\; d\epsilon \; d\alpha\ .
\end{equation}
Since $\xi^1$ and $\xi^2$ are components of a two dimensional vector,
contour deformations for them cannot be done independently because the
original contour wraps around the solutions of the scattering
equations.

We note that in the limit $\epsilon\rightarrow 0$, the dominating term
is the second summation in the second and third equations of
\eqref{deg eq}.  We can, therefore, neglect the $\rho$ dependent part,
and write the last two equations as,
\begin{equation}
  E_{n-1}^{(i)} - E_n^{(i)} = \frac{\partial}{\partial\xi^i}
  \tilde{S}^3, \qquad \text{where} \quad \tilde{S}^3 =
  \sum\limits_{a=1}^{n-2}\ese{a}{n-1}{n} \log\td{a}{\rho}{\xi}\ .
\end{equation}
Thus we have,
\begin{align}
\begin{pmatrix}
  \frac{\partial \tilde{S}^3}{\partial\xi^1}  \\
  \frac{\partial\tilde{S}^3}{\partial\xi^2}
\end{pmatrix}
=&
\begin{pmatrix}
1 & -\frac{1}{\epsilon\alpha} \\
0 & \frac{1}{\epsilon}
\end{pmatrix}
\begin{pmatrix}
  \frac{\partial \tilde{S}^3}{\partial\epsilon}  \\
  \frac{\partial\tilde{S}^3}{\partial\alpha}
\end{pmatrix} \nonumber\\
  \Rightarrow \quad \delta^{(2)}\left(\frac{\partial \tilde{S}^3}
  {\partial\xi^1}, \frac{\partial\tilde{S}^3}{\partial\xi^2}\right)
  = & \epsilon \; \delta^{(2)}\left(\frac{\partial \tilde{S}^3}
      {\partial\epsilon}, \frac{\partial\tilde{S}^3}{\partial \alpha}
      \right).
\end{align}
To see the factorisation channels we write,
\begin{equation}
  d\tilde{S}^3 = \sum\limits_{a=1}^{n-2} \ese{a}{n-1}{n}\left[\frac{d\epsilon}
    {\epsilon}
    + \frac{d\alpha}{\alpha - \alpha_{a}}\right], \qquad \alpha_{a}
  = \frac{x_{a}^2- \rho^2}{x_{a}^1- \rho^1}\ ,
\end{equation}
and find poles are at $\epsilon = 0$ and
$\alpha = \alpha_{1}, \alpha_{n-2}$. This is evident from the $\xi$
integration which follows from Eq.(\ref{deg-integral}),
\begin{align}
  & \oint\frac{\epsilon^2 d\epsilon\; d\alpha}{\frac{1}{\epsilon}
    \sum\limits_{a=1}^{n-2}\ese{a}{n-1}{n}
    \sum\limits_{a=1}^{n-2}\frac{\ese{a}{n-1}{n}}{\alpha-\alpha_{a}}}
    \left[\frac{1}{\underline{\epsilon^2}(x_{1}^{1}-\rho^1)
    (x_{n-2}^{1}-\rho^{1})\underline{(\alpha-\alpha_{1})(\alpha -
    \alpha_{n-2})}}\right]^2
    \nonumber\\
  = & \frac{1}{\sum\limits_{a=1}^{n-2}\ese{a}{n-1}{n}} \left[\frac{1}
      {\ese{1}{n-1}{n}}
      + \frac{1}{\ese{n-2}{n-1}{n}}\right]\frac{1}{\td{1}{n-2}{\rho}^2}.
\end{align}
Now we have to perform $\rho$ integration,
\begin{align}
  \mathtt{S}_{\text{deg}}^{(3)} =  \frac{1}{\sum\limits_{a=1}^{n-2}
  \ese{a}{n-1}{n}}
  & \left[\frac{1}{\ese{1}{n-1}{n}}
    + \frac{1}{\ese{n-2}{n-1}{n}}\right] \nonumber\\
  & \times\int d^2\rho\; \delta^{(2)}(E_{n-1} + E_n)\left[\frac{\td{n-3}{n-2}{1}
    \td{n-2}{1}{2}}{\td{n-3}{n-2}{\rho}\td{1}{n-2}{\rho}\td{\rho}{1}{2}}
    \right]^2.
\end{align}
The $\rho$ integration is similar to the single soft analysis with the
generalised potential function
$\tilde{S'}^{(3)} = \sum\limits_{1\leq a<b\leq n-2} (\ese{a}{b}{n-1} +
\ese{a}{b}{n}) \log \td{a}{b}{\rho} $.  We therefore obtain,
\begin{align}
\label{adg_deg_coll}
  \mathtt{S}_{\text{deg}}^{(3)}=
  & \frac{1}{\sum\limits_{a=1}^{n-2}\ese{a}{n-1}{n}}
    \Bigg(\frac{1}{\ese{n-1}{n}{1}} + \frac{1}{\ese{n-2}{n-1}{n}}
    \Bigg)\nonumber\\
  & \times \Bigg[\frac{1}{(\ese{n-3}{n-2}{n-1}+\ese{n-3}{n-2}{n})
    (\ese{n-1}{1}{2} + \ese{n}{1}{2})}\nonumber\\
  & + \frac{1}{\sum\limits_{a=1}^{n-3}(\ese{a}{n-2}{n-1}+
    \ese{a}{n-2}{n})}\left(\frac{1}{\ese{n-3}{n-2}{n-1}+\ese{n-3}{n-2}{n}}
   + \frac{1}{\ese{n-2}{n-1}{1} + \ese{n-2}{n}{1}}\right)\nonumber \\
   &  +\frac{1}{\sum\limits_{a=2}^{n-2}(\ese{a}{n-1}{1} + \ese{a}{n}{1})}
     \left(\frac{1}{\ese{n-2}{n-1}{1}+\ese{n-2}{n}{1}}
     +\frac{1}{\ese{n-1}{1}{2} + \ese{n}{1}{2}}\right) \Bigg]\ .
  \end{align}

\subsubsection{Soft punctures collinear to one hard puncture}

We will consider $\sigma_{n-1}$ and $\sigma_{n}$ to be nearly collinear
with a puncture $\sigma_{d}$ corresponding to a hard external state,
such that $|\sigma_{d}\; \sigma_{n-1}\; \sigma_{n}|\sim \tau$. It can then
be seen from Eq.(\ref{softk3}) that integrand for the double
soft factor goes as $\tau^{-2}$ whenever $d = n-2$ or $1$.  For other
values of $d$ no determinant in the denominator becomes of
$\mathcal{O}\left(\tau\right)$, therefore, this configuration produces
subleading contribution compared to the case when two soft punctures
collide.

Hence the leading double soft factor $\mathtt{S}_{\text{DS}}^{(3)}$
for the adjacent labels, $n-1$ and $n$ going soft simultaneously is
given by the expression in Eq.(\ref{adg_deg_coll}).

\subsection{Consecutive double soft limit}

We will now look at the consecutive double soft theorem, that is,
where one external particle becomes soft at a faster rate than the other
external soft particle.  Here we have two possibilities,

\paragraph{a)}
We will first take the $n$-th particle to be soft, {\em i.e.},
$s_{n\; a\; b}={\tau_{1}}{{\hat{s}}_{n\; a\; b}}$ and we know from
the single soft theorem for $k=3$ that,
\begin{align}
  m_n^{(3)}(\mathtt{I}| \mathtt{I})=\mathtt{S}^{(3)}_{n\rightarrow 0}
  m_{n-1}^{(3)}(\mathtt{I}|\mathtt{I})\ .
\end{align}
with the soft factor,
\begin{align}
  \mathtt{S}^{(3)}_{n\rightarrow 0}
  &= \frac{1}{\tau_1^2}\Bigg[\frac{1}{\hat{s}_{n-2\; n-1\; n} \hat{s}_{n\;1\;2}
    }+ \frac{1}{\sum\limits_{a=2}^{n-1} \hat{s}_{a\;n\;1}}\left(
    \frac{1}{\hat{s}_{n-1\; n\; 1}} + \frac{1}{\hat{s}_{n\;1\;2}}\right)
    \nonumber\\
  & \hspace{4.5cm}+ \frac{1}{\sum\limits_{a=1}^{n-2}\hat{s}_{a\;n-1\;n}}
    \left(\frac{1}{\hat{s}_{n-1\; n\; 1}} + \frac{1}{\hat{s}_{n-2\; n-1 \; n}}
    \right)\Bigg]\ .
\end{align}
We follow it up by considering the soft limit for the $(n-1)$-th
particle, $s_{n-1\;a\;b}=\tau_2\hat{s}_{n-1\;a\;b}$, with the
condition $\tau_{1} \ll \tau_{2}$. Therefore we obtain,
\begin{align}
  m_n^{(3)}(\mathtt{I}|\mathtt{I})=\mathtt{S}^{(3)}_{n\rightarrow 0}
  {\Bigg{|}}_{n-1\rightarrow 0} \mathtt{S}^{(3)}_{n-1\rightarrow 0}m_{n-2}^{(3)}
  (\mathtt{I}|\mathtt{I}),
\end{align}
where the second soft factor $\mathtt{S}^{(3)}_{n-1\rightarrow 0}$ has
the following form,
\begin{align}
  \mathtt{S}^{(3)}_{n-1\rightarrow 0}
  &= \frac{1}{\tau_2^2} \Bigg[\frac{1}{\hat{s}_{n-3\; n-2\; n-1}
    \hat{s}_{n-1\; 1\;2} }+ \frac{1}{\sum\limits_{a=1}^{n-3}\hat{s}_{a\;n-2\;n-1}}
    \left(\frac{1}{\hat{s}_{n-2\; n-1 \; 1}} + \frac{1}{\hat{s}_{n-3\; n-2\; n-1}}
    \right)  \nonumber\\
  & \hspace{4.5cm} + \frac{1}{\sum\limits_{a=2}^{n-2}\hat{s}_{a\;n-1\;1}}
    \left(\frac{1}{\hat{s}_{n-1\; 1\; 2}} + \frac{1}{\hat{s}_{n-2\; n-1\; 1}}
    \right)\Bigg]\ .
\end{align}
In the limit $\tau_{1} \ll \tau_{2}$, the first soft factor
$\mathtt{S}^{(3)}_{n\rightarrow 0}$ in the leading order of $\tau_1$
as well as of $\tau_2$ takes the form,
\begin{align}
  \mathtt{S}^{(3)}_{n\rightarrow 0}{\Bigg{|}}_{n-1\rightarrow 0}
  &=\frac{1}{\tau_1^2}\Bigg[\frac{1}{\tau_2\hat{s}_{n-2\; n-1\; n}
    \hat{s}_{n\;1\;2} }+ \frac{1}{\sum\limits_{a=2}^{n-1} \hat{s}_{a\;n\;1}}
    \left(\frac{1}{\tau_2\hat{s}_{n-1\; n\; 1}} + \frac{1}{\hat{s}_{n\;1\;2}}
    \right) \nonumber\\
  & \hspace{3cm} + \frac{1}{\tau_2\sum\limits_{a=1}^{n-2}\hat{s}_{a\;n-1\;n}}
    \left(\frac{1}{\tau_2\hat{s}_{n-1\; n\; 1}} + \frac{1}{\tau_2
    \hat{s}_{n-2\; n-1 \; n}}\right)\Bigg]\nonumber\\
  &= \frac{1}{\tau_1^2\tau_2^2} \left[\frac{1}{\sum\limits_{a=1}^{n-2}
    \hat{s}_{a\;n-1\;n}} \left(\frac{1}{\hat{s}_{n-1\; n\; 1}} +
    \frac{1}{\hat{s}_{n-2\; n-1 \; n}}\right)\right]\ .
\end{align}
Thus the consecutive double soft factor for the adjacent particles
where $n$-th particle is taken to be softer than the $(n-1)$-th soft
particle is given by,
\begin{align}
\label{con_soft_a}
  \mathtt{S}^{(3)}_{n\rightarrow 0}\Bigg{|}_{n-1\rightarrow 0}
  \mathtt{ S}^{(3)}_{n-1\rightarrow 0}
  &=\frac{1}{\tau_1^2\tau_2^4} \left[\frac{1}{\sum\limits_{a=1}^{n-2}
    \hat{s}_{a\;n-1\; n}} \left(\frac{1}{\hat{s}_{ n-1\; n\;1}} +
    \frac{1}{\hat{s}_{n-2\; n-1\; n}}\right)\right] \nonumber\\
  &\hspace{-5mm}\times\Bigg[ \frac{1}{\hat{s}_{n-3\; n-2\; n-1} \:
    \hat{s}_{n-1\; 1\; 2} } + \frac{1}{\sum\limits_{a=1}^{n-3}
    \hat{s}_{a\; n-2\; n-1} } \left(\frac{1}{\hat{s}_{n-3\; n-2\; n-1} }
    + \frac{1}{\hat{s}_{n-2\; n-1\; 1} }\right) \nonumber\\
  & \hspace{3.5cm} + \frac{1}{\sum\limits_{a=2}^{n-2}\hat{s}_{a\; n-1\;1} }
    \left(\frac{1}{\hat{s}_{n-2\;n-1\; 1} } + \frac{1}{\hat{s}_{n-1\; 1\; 2}}
    \right)  \Bigg]\ .
\end{align}
\paragraph{b)}
The second possibility of the consecutive adjacent double soft limit
is where we will take the $(n-1)$-th particle to be softer than the
$n$-th one, {\em i.e.}, $\tau_1 \gg \tau_2$.  The analysis is similar
to the previous one and the soft factorisation turns out to be,
\begin{equation}
  m^{(3)}_{n} \left(\mathtt{I|\mathtt{I}}\right)=
  \mathtt{S}^{(3)}_{n-1\rightarrow 0}\Bigg{|}_{n\rightarrow 0}
  \mathtt{ S}^{(3)}_{n\rightarrow 0}m^{(3)}_{n-2}
  \left(\mathtt{I|\mathtt{I}}\right)\ ,
\end{equation}
where the soft factor is given by,
\begin{align}
\label{con_soft_b}
  \mathtt{S}^{(3)}_{n\rightarrow 0}\Bigg{|}_{n-1\rightarrow 0}
  \mathtt{ S}^{(3)}_{n-1\rightarrow 0}
  &=\frac{1}{\tau_1^4\tau_2^2}\left[\frac{1}{\sum\limits_{a=1}^{n-2}
    \hat{s}_{a\;n-1\; n}} \left(\frac{1}{\hat{s}_{ n-1\; n\;1}} +
    \frac{1}{\hat{s}_{n-2\; n-1\; n}}\right)\right] \nonumber\\
  & \times\Biggl[ \frac{1}{\hat{s}_{n-3\; n-2\; n} \hat{s}_{n\;1\;2} }+
    \frac{1}{\sum\limits_{a=1}^{n-3}\hat{s}_{a\;n-2\;n}}
    \left(\frac{1}{\hat{s}_{n-3\; n-2\; n}} +
    \frac{1}{\hat{s}_{n-2\; n\; 1}}\right)\nonumber\\
  & \hspace{3.2cm} + \frac{1}{\sum\limits_{a=2}^{n-2} \hat{s}_{a\;n\;1}}
    \left(\frac{1}{\hat{s}_{n-2\; n\; 1}} + \frac{1}{\hat{s}_{n\;1\;2}}\right)
    \Biggr]\ .
\end{align}
We can obtain Eq.(\ref{con_soft_a}) and Eq.(\ref{con_soft_b}) by
taking appropriate consecutive limits in $\tau_1$ and $\tau_2$
starting from the expression of simultaneous double soft limit derived
in Eq.(\ref{adg_deg_coll}). This serves as a consistency check for the
simultaneous double soft factor.


\section{Double Soft Theorem for Arbitrary $k$}
\label{sec:gen-k-adjacent}

The result of sec:(\ref{sec:APS}) can be generalised for arbitrary
$k$. Contributions for the leading simultaneous double soft factor
when two adjacent particles are taken to be soft come from degenerate
solutions, more specifically when the two corresponding soft punctures
are infinitesimally close to each other, {\em i.e.}, when the
separation is of $\mathcal{O}\left(\tau\right)$. The other degenerate
configuration occurs when one of the soft punctures approaches the
co-dimension one subspace generated by other soft puncture and $(k-2)$
number of hard punctures.  This degeneration, however, contributes at
subleading order in the soft theorem. Needless to say, the
non-degenerate solutions appear at further subleading orders in the
expansion and therefore we do not discuss them here.  In this section,
we present only the leading order result for the double soft theorem
in adjacent simultaneous soft limit for any $k \ge 3$.

We will now consider the soft limit in labels $n-1$ and $n$.  We can
then impose following conditions on the generalised Mandelstam variables:
\begin{equation}
\begin{aligned}
s_{n-1\; a_{1} \cdots a_{k-1}} & = \tau \hat{s}_{n-1\; a_{1} \cdots a_{k-1}}\ ,\\
s_{n\; a_{1} \cdots a_{k-1}} & = \tau \hat{s}_{n\; a_{1} \cdots a_{k-1}}\ ,\\
s_{n-1\; n\; a_{1}\cdots a_{k-2}} & = \tau^{2} \hat{s}_{n-1\; n\; a_{1}\cdots a_{k-2}}\ ,
\end{aligned}
\end{equation}
for $a_{1} ,\cdots a_{k-1} \in \{1,2, \cdots n-2\}$ labelling the hard
external states, and rest of the Mandelstam variables are of order
unity.

We are interested in the soft limit
$m_{n}^{(k)} \left(\textsc{I}|\textsc{I}\right) =
\mathtt{S}_{\text{DS}}^{(k)} \;m_{n-2}^{(k)}
\left(\textsc{I}|\textsc{I}\right)$ for $k\ge 3$, where the double soft
factor can be expressed in the integral form as,
\begin{eqnarray}\label{eq:gen-k-soft1}
  \mathtt{S}_{\text{DS}}^{(k)}
  & = & \int d^{k-1}x_{n} \; \delta^{(k-1)}\left(E_{n}\right)
        \int d^{k-1}x_{n-1}\; \delta^{(k-1)}\left(E_{n-1}\right)\times \nonumber\\
  && \hspace{-1.5cm}\left[\frac{|n-k \cdots 1 | |(n-k+1) \cdots 1\; 2|
     \cdots |n-2\; 1\cdots k-1|}{|n-k \cdots n-2\, \sigma_{n-1}|
     |(n-k+1)\cdots \sigma_{n-1}\,\sigma_{n}|\cdots |
     \sigma_{n-1}\,\sigma_{n}\,1\cdots k-2||\sigma_{n}\, 1\cdots k-1|}
     \right]^{2}\ . \nonumber\\
\end{eqnarray}
In the soft limit we can write the scattering equations as follows:
\begin{eqnarray}\label{eq:gen-k-scatt1}
  E_{a_{1}}^{(i)}
  & = & \sum\limits_{1\le a_{2} \cdots < a_{k} \le n-2}
        \frac{s_{a_{1}\cdots a_{k}}}{|a_{1}\cdots a_{k}|}
        \frac{\partial}{\partial x_{a_{1}}^{i}}|a_{1}\cdots a_{k}| = 0,
        \qquad \forall a_{1}  \in \{1,2, \cdots n-2\} \nonumber\\
  E_{n-1}^{(i)}
  & = & \sum\limits_{1\le a_{1} \cdots < a_{k-1} \le n-2}
        \frac{s_{a_{1}\cdots a_{k-1}\; n-1}}{|a_{1}\cdots a_{k-1}\;
        \sigma_{n-1}|}\frac{\partial}{\partial x_{n-1}^{i}}|
        a_{1}\cdots a_{k-1}\; \sigma_{n-1}|\nonumber\\
  && \hspace{1cm} + \sum\limits_{1\le a_{1} \cdots < a_{k-2}\le n-2}
     \frac{s_{a_{1}\cdots a_{k-2}\; n-1\; n}}{|a_{1}\cdots a_{k-2}\;
     \sigma_{n-1}\; \sigma_{n}|}\frac{\partial}{\partial x_{n-1}^{i}}
     |a_{1}\cdots a_{k-2}\; \sigma_{n-1}\; \sigma_{n}| = 0\ ,\nonumber\\
  E_{n}^{(i)}
  & = & \sum\limits_{1\le a_{1} \cdots < a_{k-1} \le n-2}
        \frac{s_{a_{1}\cdots a_{k-1}\; n}}{|a_{1}\cdots a_{k-1}\; \sigma_{n}|}
        \frac{\partial}{\partial x_{n}^{i}}|a_{1}\cdots a_{k-1}\; \sigma_{n}|
        \nonumber\\
  && \hspace{1cm} + \sum\limits_{1\le a_{1} \cdots < a_{k-2}\le n-2}
     \frac{s_{a_{1}\cdots a_{k-2}\; n-1\; n}}{|a_{1}\cdots a_{k-2}\;
     \sigma_{n-1}\; \sigma_{n}|}\frac{\partial}{\partial x_{n}^{i}}
     |a_{1}\cdots a_{k-2}\; \sigma_{n-1}\; \sigma_{n}| = 0\ ,
\end{eqnarray}
where $i = 1,2, \cdots k-1$.  We would again like to emphasise that
here we are only considering the regular solutions and hence have
ignored any additional $\mathcal{O}\left(\tau^{0}\right)$ terms in
$E_{a}^{(i)}$ which will be present if we include the singular
solutions.  To be precise we assume the condition that
$|a_{1}\cdots a_{k-1} \sigma_{n-1}|$ and
$|a_{1}\cdots a_{k-1} \sigma_{n}|$ are always of order one.

We will now focus on the degenerate solutions, {\em i.e.}, when
$|a_{1}\,\cdots\, a_{k-2}\, \sigma_{n-1}\,\sigma_{n}|$ is of
$\mathcal{O}(\tau)$.  There are two way to arrive this condition, when
$\sigma_{n-1}$ and $\sigma_{n}$ are in $\mathcal{O}(\tau)$
neighbourhood of each other or when the two soft punctures and any set
of $(k-2)$ hard punctures $a_{j}, \ j \in \{1,2,\cdots n-2\}$ form a
co-dimension one subspace up to $\mathcal{O}(\tau)$ deformation.
However, following our analysis earlier, we will ignore the latter
configuration.

We choose the following change of variables,
\begin{eqnarray}
x_{n-1}^{i} &=& \rho^{i} + \xi^{i}\ , \nonumber\\
  x_{n}^{i} & = & \rho^{i} - \xi^{i}\ , \qquad \xi^{i} \sim \mathcal{O}
                  \left(\tau\right), \quad i = 1, \cdots k-1\ .
\end{eqnarray}
The integration measure transforms in the following way,
\begin{equation}
d^{k-1}x_{n-1}\; d^{k-1}x_{n} = 2^{k-1}d^{k-1}\rho\; d^{k-1}\xi\ ,
\end{equation}
and the delta functions with arguments as $E_{n}$ and $E_{n-1}$ are
expressed as,
\begin{equation}
  \delta^{(k-1)}\left(E_{n}\right)\; \delta^{(k-1)}\left(E_{n-1}\right) =
  2^{k-1}\delta^{(k-1)}\left(E_{n-1}+E_{n}\right)\; \delta^{(k-1)}
  \left(E_{n-1}-E_{n}\right).
\end{equation}
For later purpose we define,
\begin{eqnarray}
|a_{1}\;\cdots a_{k-2}\; \sigma_{n-1}\; \sigma_{n}| & = & - 2\begin{vmatrix}
1 & 1& \cdots & 1& 1& 0\\
x_{a_{1}}^{1} & x_{a_{2}}^{1} & \cdots & x_{a_{k-2}}^{1} & \rho^{1} & \xi^{1}\\
\vdots & \vdots & \vdots & \vdots & \vdots & \vdots \\
x_{a_{1}}^{k-1} & x_{a_{2}}^{k-1} & \cdots & x_{a_{k-2}}^{k-1} &
\rho^{k-1} & \xi^{k-1}
\end{vmatrix} \nonumber\\
& : = & -2 \Delta^{(k)}_{a_{1}\,\cdots\, a_{k-2}\,\rho\,\xi}.
\end{eqnarray}
The soft factor in Eq.(\ref{eq:gen-k-soft1}) then becomes,
\begin{eqnarray}\label{eq:gen-k-soft2}
  \mathtt{S}_{\text{DS}}^{(k)}
  & = & \int d^{k-1}\rho \; d^{k-1}\xi \; \delta^{(k-1)}
        \left(E_{n-1}+E_{n}\right)
        \; \delta^{(k-1)}\left(E_{n-1}-E_{n}\right) \nonumber\\
  && \times \left[\frac{|n-k \cdots 1 | |(n-k+1) \cdots 1\; 2|
     \cdots |n-2\; 1\cdots k-1|}{\Delta^{(k)}_{(n-k+1)\,\cdots\, n-2\,\rho\,\xi}
     \cdots \Delta^{(k)}_{1\,\cdots\, k-2\,\rho\,\xi} |n-k\;\cdots n-2\;\rho|
     |\rho\; 1\;\cdots\;k-1|}\right]^{2}.
\end{eqnarray}
The last two equations in (\ref{eq:gen-k-scatt1}) can be expressed as,
\begin{eqnarray}\label{eq:gen-k-scatt2}
  E_{n-1}^{(i)}
  & = & \sum\limits_{1\le a_{1}\cdots <a_{k-1}\le n-2}
        \frac{s_{a_{1}\cdots a_{k-1}\; n-1}}{|a_{1}\cdots a_{k-1}\; \rho|}
        \frac{\partial}{\partial\rho^{i}}|a_{1}\cdots a_{k-1}\; \rho|
        \nonumber\\
  && \hspace{1cm} + \sum\limits_{1\le a_{1}\cdots < a_{k-2}\le n-2}
     \frac{s_{a_{1}\cdots a_{k-2}\; n-1\; n}}
     {2\Delta_{a_{1}\,\cdots\, a_{k-2}\,\rho\,\xi}^{(k)}}
     \frac{\partial}{\partial\xi^{i}}
     \Delta_{a_{1}\,\cdots\, a_{k-2}\,\rho\,\xi}^{(k)} = 0, \nonumber\\
  E_{n}^{(i)}
  & = & \sum\limits_{1\le a_{1}\cdots <a_{k-1}\le n-2}
        \frac{s_{a_{1}\cdots a_{k-1}\; n}}{|a_{1}\cdots a_{k-1}\; \rho|}
        \frac{\partial}{\partial\rho^{i}}|a_{1}\cdots a_{k-1}\; \rho|
        \nonumber\\
  && \hspace{1cm} - \sum\limits_{1\le a_{1}\cdots < a_{k-2}\le n-2}
     \frac{s_{a_{1}\cdots a_{k-2}\; n-1\; n}}
     {2\Delta_{a_{1}\,\cdots\, a_{k-2}\,\rho\,\xi}^{(k)}}\frac{\partial}
     {\partial\xi^{i}}\Delta_{a_{1}\,\cdots\, a_{k-2}\,\rho\,\xi}^{(k)} = 0.
\end{eqnarray}
We first do the $\xi$ integrals in Eq.(\ref{eq:gen-k-soft2}), for which we
make a change of variables,
\begin{eqnarray}
\xi^{1} & = & \epsilon\ , \nonumber\\
\xi^{j} & = & \epsilon\zeta^{j}\ , \quad j = 2,\cdots k-1\ .
\end{eqnarray}
This implies,
\begin{equation}
d^{k-1}\xi = \epsilon^{k-2}d\epsilon\; d^{k-2}\zeta\ .
\end{equation}
Following the above change of variables we can write
$\Delta^{(k)}_{a_{1}\,\cdots\, a_{k-2}\,\rho\,\xi}$ as,
\begin{eqnarray}
  \Delta^{(k)}_{a_{1}\,\cdots\, a_{k-2}\,\rho\,\xi}
  & = & \epsilon\left(x_{a_{1}}^{1} - \rho^{1}\right)\cdots
        \left(x_{a_{k-2}}^{1} - \rho^{1}\right)
        \Delta^{(k-1)}_{\zeta\; a_{1}\,\cdots\, a_{k-2}},
\end{eqnarray}
where we have defined,
\begin{equation}\label{deta-def}
 \Delta^{(k-1)}_{\zeta\; a_{1}\;\cdots\; a_{k-2}} := \begin{vmatrix}
1 & 1 & \cdots & 1  \\
\zeta^{2} & \alpha_{a_{1}}^{2} & \cdots & \alpha_{a_{k-2}}^{2} \\
\vdots & \vdots & \vdots & \vdots \\
\zeta^{k-1} & \alpha_{a_{1}}^{k-1} & \cdots & \alpha_{a_{k-2}}^{k-1}
\end{vmatrix}\ , \qquad \alpha_{a}^{j} = \frac{x_{a}^{j} - \rho^{j}}
{x_{a}^{1} - \rho^{1}}\ .
\end{equation}
Similarly $\Delta^{(k-1)}_{a_{1}\;\cdots\; a_{k-2}\; a_{k-1}}$ will be
denoted by the determinant of $(k-1)\times (k-1)$ minor obtained by
replacing the column
$\begin{pmatrix} 1 \\ \zeta^{2} \\ \vdots \\ \zeta^{k-1}
\end{pmatrix}$ by $\begin{pmatrix}
1 \\ \alpha_{k-1}^{2} \\ \vdots \\ \alpha_{k-1}^{k-1} 
\end{pmatrix}$ in Eq.(\ref{deta-def}).

We define a new potential function,
\begin{equation}
  \tilde{S} = \sum\limits_{1\le a_{1} \cdots < a_{k-2}\le n-2}
  s_{a_{1}\cdots a_{k-2}\; n-1\; n}\log
  \Delta^{(k)}_{a_{1}\,\cdots\, a_{k-2}\,\rho\,\xi}\ ,
\end{equation}
such that near $\epsilon\to 0$, $E_{n-1}^{(i)} - E_{n}^{(i)}$
can be approximated to be,
\begin{equation}
  E_{n-1}^{(i)} - E_{n}^{(i)} \approx \sum\limits_{1\le a_{1}\cdots < a_{k-2}\le n-2}
  \frac{s_{a_{1}\cdots a_{k-2}\; n-1\; n}}
  {\Delta_{a_{1}\,\cdots\, a_{k-2}\,\rho\,\xi}^{(k)}}
  \frac{\partial}{\partial\xi^{i}}\Delta_{a_{1}\,\cdots a_{k-2}\,\rho\,\xi}^{(k)}
  = \frac{\partial}{\partial\xi^{i}}\tilde{S} = 0\ .
\end{equation}
The delta function with the argument $E_{n-1}^{(i)} - E_{n}^{(i)}$ can be used
to perform $\xi$ integration as a contour integral.  Under the change of
coordinates from $\xi^{i}$ to $\left(\epsilon, \zeta^{j}\right)$ this
delta function transforms as,
\begin{equation}
  \delta^{(k-1)}\left(E_{n-1}-E_{n}\right) = \epsilon^{k-2}
  \delta\left(\frac{\partial \tilde{S}}{\partial\epsilon}\right)
  \; \delta^{(k-2)}\left(\frac{\partial \tilde{S}}{\partial\zeta^{j}}\right).
\end{equation}
$\epsilon$ integration gives a residue
$\left(\sum\limits_{1\le a_{1}\cdots <a_{k-2}\le n-2}s_{a_{1}\cdots
    a_{k-2}\; n-1\; n}\right)^{-1}$. The soft factor in
Eq.(\ref{eq:gen-k-soft2}) now takes the form,
\begin{eqnarray}\label{eq:gen-k-soft-final}
  \mathtt{S}_{\text{DS}}^{(k)}
  & = & \frac{1}{\sum\limits_{1\le a_{1}\cdots <a_{k-2}\le n-2}
        s_{a_{1}\cdots a_{k-2}\; n-1\; n}}\bigintss d^{k-2}\zeta\;
        \delta^{(k-2)}\left(\frac{\partial \tilde{S}}
        {\partial\zeta^{j}}\right)\nonumber\\
  && \hspace{0.5cm} \times \left[\frac{\Delta^{(k-1)}_{(n-k+1)\cdots 1}
     \cdots \Delta^{(k-1)}_{n-2\; 1\cdots k-2}}
     {\Delta^{(k-1)}_{(n-k+1)\cdots n-2\;\zeta} \cdots
     \Delta^{(k-1)}_{n-2\; \zeta\cdots 1} \Delta^{(k-1)}_{\zeta\;1\cdots k-2}}
     \right]^{2}\bigintss d^{k-1}\rho \; \delta^{(k-1)}\left(E_{n-1}+E_{n}
     \right) \nonumber\\
  && \hspace{0.5cm}\times  \Biggl[\frac{|n-k \cdots 1 |
     |n-\left(k-1\right) \cdots 1\; 2| \cdots |n-2\; 1\cdots k-1|}
     { |n-k \cdots n-2\; \rho| |(n-k+1)\cdots \rho\; 1|\cdots|\rho\;
     1\cdots k-1|}\Biggr]^{2} \nonumber\\
&&\nonumber\\
  & = & \frac{1}{\sum\limits_{1\le a_{1}\cdots <a_{k-2}\le n-2}
        s_{a_{1}\cdots a_{k-2}\; n-1\; n}} \;\mathtt{S}^{(k-1)}
        \left(s_{a_{1}\cdots a_{k-2}\; m} \rightarrow
        s_{a_{1}\cdots a_{k-2}\; n-1\; n} \right) \; \mathtt{S}^{(k)}.
\end{eqnarray}
This is the expression for double soft factor for arbitrary $k$ when
adjacent particles are taken to be soft. $\mathtt{S}^{(k-1)}$ is the
single soft factor corresponding to $k-1$ which has the pole structure
of the form $s_{a_{1}\cdots a_{k-2}\; n-1\; n}$ in place of usual
$(k-1)$-indexed Mandelstam variables. $\mathtt{S}^{(k)}$ denotes the
single soft factor for $k$ with shifted propagators of the form
$s_{a_{1}\cdots a_{k-1}\; n-1} + s_{a_{1}\cdots a_{k-1}\;
  n}$. Eq.(\ref{eq:gen-k-soft-final}) provides a recursion relation of
calculating double soft factor from single soft factors.  It is worth
pointing out that the leading contribution to the double soft factor
for arbitrary $k$ scales at $\tau^{-3(k-1)}$.


\section{Next to Adjacent Soft Limits for $k=3$ Amplitudes}
\label{sec:nasl3a}

In this section, we study the double soft limits for next to adjacent
external states labelled by $n-2$ and $n$. For this, we choose the
scaling of the generalised Mandelstam variables as,
\begin{equation}
  \ese{a}{b}{n-2} =\tau\hat{s}_{a\,b\,n-2}, \qquad \ese{a}{b}{n} =
  \tau\hat{s}_{a\,b\,n},
  \qquad \ese{a}{n-2}{n} = \tau^2\hat{s}_{a\,n-2\,n}, \qquad
  a, b \ne n-2, n\ ,
\end{equation}
where, $a,b$ label the hard punctures.  We will consider regular
solutions and ignore the singular solutions where the determinants
$|a\;b\;n|$, and $|a\;b\;n-2|$ scale as $\tau$, the scattering
equations become,
\begin{align}\label{scatt_eqn_non-adj}
  E_a^{(i)}
  = & \sum\limits_{b,c \ne a, n-2, n} \frac{\ese{a}{b}{c}}{\td{a}{b}{c}}
      \frac{\partial}{\partial x_a^{(i)}}\td{a}{b}{c} = 0\ , \quad
      \forall a\ , \nonumber\\
  E_{n-2}^{(i)}
  = & \tau \sum\limits_{a,b\ne n-2, n}\frac{\hat{s}_{a\,b\,n-2}}
      {\td{a}{b}{n-2}}\frac{\partial}{\partial x_{n-2}^{(i)}}
      \td{a}{b}{n-2} \nonumber\\
      & \hspace{5mm}+\tau^2 \sum\limits_{a\ne n-2, n}
      \frac{\hat{s}_{a\,n-2\,n}}{\td{a}{n-2}{n}}
      \frac{\partial}{\partial x_{n-2}^{(i)}}\td{a}{n-2}{n} =0\ ,
      \\
  E_n^{(i)}
  = & \tau \sum\limits_{a,b\ne n-2, n}\frac{\hat{s}_{a\,b\,n}}
      {\td{a}{b}{n}}\frac{\partial}{\partial x_n^{(i)}}\td{a}{b}{n}\nonumber\\
      & \hspace{5mm}+ \tau^2\sum\limits_{a\ne n-2,n}\frac{\hat{s}_{a\,n-2\,n}}
      {\td{a}{n-2}{n}}\frac{\partial}{\partial x_n^{(i)}}
      \td{a}{n-2}{n} = 0\nonumber\ .
\end{align}
The soft factor for the next to adjacent soft punctures can then be
expressed as,
\begin{align}
  \mathtt{S}^{(3)}_{n, n-2}
  & = \int d^2 x_n\int d^2 x_{n-2}\; \delta^{(2)}(E_n)
      \delta^{(2)}(E_{n-2}) \nonumber\\
    &\times \left[\frac{\td{n-4}{n-3}{n-1}\td{n-3}{n-1}{1}
       \td{n-1}{1}{2}}{\td{n-4}{n-3}{n-2}\td{n-3}{n-2}{n-1}\td{n-2}{n-1}{n}
       \td{n-1}{n}{1}\td{n}{1}{2}}\right]^2\ .
\end{align}
In the next subsection, we will study the consecutive double soft limit
to understand the leading order behaviour of the soft factor and later
we will take up the simultaneous double soft limit as a consistency
check on the results of the consecutive soft limit.
%
\subsection{Consecutive double soft limit}

In the consecutive double soft limit, we will encounter two
possibilities depending on which external state becomes soft at a
faster rate.
\paragraph{{a)}}
We will start with the case where the $n$-th particle is softer than
the $(n-2)$-th one where $s_{n\;a\;b}=\tau_1\hat{s}_{n\;a\;b}$ and
$s_{n-2\;a\;b}=\tau_2\hat{s}_{n-2\;a\;b}$, with
{$\tau_{1}\ll\tau_{2}$}.  After taking these two limits we have,
\begin{align}
  m_n^{(3)}(\mathtt{I}|\mathtt{I})=\mathtt{S}^{(3)}_{n\rightarrow 0}
  \Bigg|_{n-2\rightarrow 0}\mathtt{S}^{(3)}_{n-2\rightarrow 0}m_{n-2}^{(3)}
  (\mathtt{I}|\mathtt{I}),
\end{align} 
where,
\begin{eqnarray}
  \mathtt{S}^{(3)}_{n-2\rightarrow 0}
  &=& \frac{1}{\tau_2^2} \Biggl[\frac{1}{\hat{s}_{n-4\; n-3\; n-2}\
      \hat{s}_{n-2\; n-1\;1} }+ \frac{1}{\sum\limits_{a=1}^{n-3}
      \hat{s}_{n-2\;a\;n-1}}\left(\frac{1}{\hat{s}_{n-3\; n-2\; n-1}} +
      \frac{1}{\hat{s}_{n-2\; n-1\;1}}\right)\nonumber\\
  &&+\frac{1}{\sum\limits_{a=1,a\neq n-2, n-3}^{n-1}\hat{s}_{n-2\;a\;n-3}}
     \left(\frac{1}{\hat{s}_{n-3\; n-2 \; n-1}} + \frac{1}
     {\hat{s}_{n-4\; n-3\; n-2}}\right)\Biggr]\ ,
 \end{eqnarray}
and,
\begin{align}
  \mathtt{S}^{(3)}_{n\rightarrow 0}\Bigg|_{n-2\rightarrow 0}
  &=\frac{1}{\tau_1^2}\Biggl\{\frac{1}{\tau_2\hat{s}_{n-2\; n-1\; n}\
    \hat{s}_{n\;1\;2} }+ \frac{1}{\sum\limits_{a=2}^{n-1} \hat{s}_{1\;a\;n}}
    \left(\frac{1}{\hat{s}_{n-1\; n\; 1}} + \frac{1}{\hat{s}_{n\;1\;2}}\right)
    \nonumber\\
  &\hspace{3cm}+ \frac{1}{\sum\limits_{a=1}^{n-2}\hat{s}_{a\;n-1\;n}}
    \left(\frac{1}{\tau_2\hat{s}_{n-2\; n-1 \; n}} +
    \frac{1}{\hat{s}_{n-1\; n\; 1}}\right)\Biggr\}\nonumber\\
  &= \frac{1}{\tau_1^2\tau_2} \left\{\frac{1}{\hat{s}_{n-2\; n-1\; n}\
    \hat{s}_{n\;1\;2} }+\frac{1}{\sum\limits_{a=1}^{n-3}\hat{s}_{a\;n-1\;n}}
    \frac{1}{\hat{s}_{n-2\; n-1 \; n}}\right\}.
\end{align}
The full soft factor for this consecutive limit with
{$\tau_{1}\ll\tau_{2}$} is,
\begin{eqnarray}\label{next_adj_consecutive_a}
  \mathtt{S}^{(3)}_{n\rightarrow 0}\Bigg|_{n-2\rightarrow 0}
  \mathtt{S}^{(3)}_{n-2\rightarrow 0}
  & = & \frac{1}{\tau_1^2\tau_2^3}
        \Biggl\{\frac{1}{\hat{s}_{n-2\; n-1\; n}} \Bigg( \frac{1}
        {\hat{s}_{n\;1\;2}}+\frac{1}{\sum\limits_{a=1}^{n-3}
        \hat{s}_{a\;n-1\;n}} \Bigg)\Biggr\}
        \Bigg[\frac{1}{\hat{s}_{n-4\; n-3\; n-2}\ \hat{s}_{n-2\; n-1\;1} }
        \nonumber\\
  &&+ \frac{1}{\sum\limits_{a=1,a\neq n-2, n-1}^{n-3}\hat{s}_{n-2\;a\;n-1}}
     \left(\frac{1}{\hat{s}_{n-3\; n-2\; n-1}} + \frac{1}{\hat{s}_{n-2\; n-1\;1}}
     \right)\nonumber\\
  && +  \frac{1}{\sum\limits_{a=1,a\neq n-2, n-3}^{n-1}\hat{s}_{n-2\;a\;n-3}}
     \left(\frac{1}{\hat{s}_{n-3\; n-2 \; n-1}} +
     \frac{1}{\hat{s}_{n-4\; n-3\; n-2}}\right)\Bigg]\ .
\end{eqnarray}
\paragraph{{b)}} If instead we take {$\tau_{2}\ll\tau_{1}$} and carry
out a similar analysis as in the previous case, the consecutive
soft factor becomes,
\begin{eqnarray}\label{next_adj_consecutive_b}
  \mathtt{S}^{(3)}_{n-2\rightarrow 0}\Bigg|_{ n\rightarrow 0}
  \mathtt{S}^{(3)}_{n\rightarrow 0}
  & = &  \frac{1}{\tau_1^3\tau_2^2}\Biggl\{ \frac{1}{\hat{s}_{n-2\; n-1\; n}}
    \Bigg( \frac{1}{\hat{s}_{n-4\;n-3\;n-2} }+\frac{1}{\sum\limits_{a=1}^{n-3}
        \hat{s}_{n-2\;a\;n-1}} \Bigg)\Biggr\} \nonumber\\
  &&\hspace{1cm} \times \Bigg[\frac{1}{\hat{s}_{n-3\; n-1\; n}\ \hat{s}_{n\;1\;2}}
    + \frac{1}{\sum\limits_{a=2}^{n-1} \hat{s}_{1\;a\;n}}
     \left(\frac{1}{\hat{s}_{n-1\; n\; 1}} + \frac{1}{\hat{s}_{n\;1\;2}}\right)
     \nonumber\\
  && \hspace{3cm} + \frac{1}{\sum\limits_{a=1}^{n-3}\hat{s}_{a\;n-1\;n}}
     \left(\frac{1}{\hat{s}_{n-3\; n-1 \; n}} + \frac{1}{\hat{s}_{n-1\; n\; 1}}
     \right)\Bigg]\ .
\end{eqnarray}
It is now evident from \eq{next_adj_consecutive_a} and
\eq{next_adj_consecutive_b} that the leading behaviour of the double
soft factor in the simultaneous limit is expected to be
$\mathcal{O}(\tau^{-5})$.  On the other hand, the scaling of the soft
factor for the non-degenerate solutions goes as
$\mathcal{O}(\tau^{-4})$, which is subleading compared to the
degenerate case.  This suggests that if we want to pick up the leading
effects in the simultaneous limit, it suffices to look at the
degenerate solutions.  However, in appendix \ref{sec:app1}, we have
analysed the contribution of non-degenerate solutions to the subleading
results.

\subsection{Simultaneous double soft limit: Degenerate solutions}

We will now look at the simultaneous double soft limit.  In this case
the determinant $\td{a}{n-2}{n}$ scales as $\tau$, for $a$ belonging
to hard particles.  This scaling happens when either the two soft
punctures come close to each other, or they are nearly collinear with
one of the hard punctures.  We will look at both these possibilities
now.

\subsubsection{Collision configuration}

To study the colliding configurations we consider the following change
of variables,
\begin{equation}
x_{n-2}^i= \rho^i+\xi^i, \qquad x_n^i= \rho^i-\xi^i, \qquad i=1,2\ ,
\end{equation}
where $\xi$ is $\mathcal{O}(\tau)$. In terms of these new variables,
scattering equations become,
\begin{align}
  E_{n-2}^{(i)} + E_n^{(i)}
  = & \sum\limits_{a,b} \frac{\ese{a}{b}{n-2} + \ese{a}{b}{n}}
      {\td{a}{b}{\rho}}\frac{\partial}{\partial\rho^i} \td{a}{b}{\rho}\ ,
      \nonumber\\
  E_{n-2}^{(i)} - E_n^{(i)}
  = & \sum\limits_{a,b} \frac{\ese{a}{b}{n-2} - \ese{a}{b}{n}}{\td{a}{b}{\rho}}
      \frac{\partial}{\partial\rho^i} \td{a}{b}{\rho} + \sum\limits_a
      \frac{\ese{a}{n-2}{n}}{\td{a}{\rho}{\xi}}\frac{\partial}{\partial\xi^i}
      \td{a}{\rho}{\xi}\ .
\end{align}
The soft factor for the degenerate solutions can be expressed as,
\begin{align}
  \mathtt{S}^{(3)}_{\text{deg}}
  = &  \int d^2\rho\; d^2\xi \; \delta^{(2)}(E_{n-2} + E_n)
      \delta^{(2)}(E_{n-2} - E_n) \nonumber\\
    & \hspace{1cm}\times \left[\frac{\td{n-4}{n-3}{n-1}\td{n-3}{n-1}{1}
      \td{n-1}{1}{2}}{\td{n-4}{n-3}{\rho}\td{n-3}{\rho}{n-1}
      \td{\rho}{n-1}{\xi}\td{n-1}{\rho}{1}\td{\rho}{1}{2}}\right]^2\ .
\end{align}
We can first do the $\xi$ integration by deforming the contour.  To do
that we will consider following parametrisation,
\begin{align}
\xi^1 =&  \epsilon\ , \nonumber\\
\xi^2 =&  \epsilon\alpha\ .
\end{align}
The integration measure can be written as,
\begin{equation}
d^2\xi = \epsilon\; d\epsilon \; d\alpha.
\end{equation}
Similarly delta functions will transform with a Jacobian factor, which
is equal to $\epsilon$.  We can then express the $\xi$ integral in
terms of $\epsilon$ and $\alpha$.  It is easy to see that there is no
singularity in the $\epsilon$ contour integral and as a result this
integral vanishes,
\begin{align}
  & \oint \frac{\epsilon^2d\epsilon \; d\alpha}{\frac{1}{\epsilon}
    \sum\limits_a\ese{a}{n-2}{n}
     \sum\limits_a\frac{\ese{a}{n-2}{n}}
    {\alpha-\alpha_a}}\left[\frac{1}{\underline{\epsilon
    (\rho^1-x_{n-1}^1)(\alpha-\alpha_{n-1})}}\right]^2, \qquad
    \alpha_a = \frac{x_a^2-\rho^2}{x_a^1-\rho^1} \nonumber\\
  = & \frac{1}{\sum\limits_a \ese{a}{n-2}{n}}\oint \epsilon\; d\epsilon
      \oint \frac{d\alpha}{\sum\limits_a\frac{\ese{a}{n-2}{n}}
      {\alpha-\alpha_a}}\left[\frac{1}{\underline{
      (\rho^1-x_{n-1}^1)(\alpha-\alpha_{n-1})}}\right]^2\nonumber\\
  = & 0\ .
\end{align}
In fact, the collision singularity contributes at the order
$\tau^{-4}$, and in that sense, this vanishing contribution is
subleading.  This hints that the dominating contribution comes from
the collinear singularity, where two soft punctures become collinear
with one hard puncture.

\subsubsection{Soft punctures collinear to one hard puncture}

We will now consider the case when the determinant $\td{{n-2}}{n}{d}$
scales as $\tau$, {\em i.e.}, both the soft punctures are nearly
collinear to one of the hard punctures labelled by $d$.  We will now
use the collinearity ansatz by parametrising $x_{n-2}^i$ as,
\begin{align}
x_{n-2}&=x_d+\alpha (x_n-x_d)\ ,\nonumber\\	
y_{n-2}&=y_d+\alpha (y_n-y_d)+\tau\xi\ .
\end{align}
Therefore the determinant $\td{n-2}{n}{d}$ becomes,
\begin{align}
\td{n-2}{n}{d}&=\tau\xi(x_d-x_n)\ .
\end{align}
The Parke-Taylor contribution for $d = n-1$ then can be written as,
\begin{align}
  \left[\frac{\td{n-4}{n-3}{n-1}\td{n-3}{n-1}{1}\td{n-1}{1}{2}}{\tau
  \xi(x_{n-1} - x_n)
  \td{n-4}{n-3}{\alpha(\sigma_n-\sigma_{n-1})}
  \td{n-3}{\alpha(\sigma_n-\sigma_{n-1})}{n-1}\td{n-1}{n}{1}
  \td{n}{1}{2}}
  \right]^2,
\end{align}
where we have used the definition,
\begin{align}
\td{a}{b}{\alpha (\sigma_n-\sigma_{n-1)}}&=\td{a}{b}{d}+\begin{vmatrix}
1 & 1 & 0 \\
x_a & x_b & \alpha(x_n-x_{n-1})\\
y_a & y_b & \alpha(y_n-y_{n-1})
\end{vmatrix}\ .
\end{align}
For $d\ne n-1$, the Parke-Taylor contribution is ${\cal O}(\tau^0)$.
For $d=n-1$, the measure becomes,
\begin{align}
dx_ndy_ndx_{n-2}dy_{n-2}=\tau(x_n-x_{n-1})dx_ndy_nd\alpha d\xi.
\end{align}
The scattering equations for $n$ and $n-2$ become, 
\begin{align}
  E_{n-2}^{(1)}
  & = \tau \sum\limits_{1\leq a< b\leq n-1, a,b \ne n-2}\frac{\hat{s}_{a\;b\;n-2}}
    {\td{a}{b}{\alpha (\sigma_n-\sigma_{n-1)}}}\left(y_a- y_b\right) - \tau
    \frac{\hat{s}_{n-2\;n-1\;n}}{\xi(x_n-x_{n-1})}(y_n-y_{n-1}), \nonumber\\
  E_{n-2}^{(2)}
  & = -\tau \sum\limits_{1\leq a< b\leq n-1, a,b \ne n-2} \frac{\hat{s}_{a\;b\;n-2}}
    {\td{a}{b}{\alpha (\sigma_n-\sigma_{n-1)}}}\left(x_a- x_b\right) +
    \tau \frac{\hat{s}_{n-2\;n-1\;n}}{\xi}, \nonumber\\
  E_{n}^{(1)}
  & = \tau \sum\limits_{1\leq a< b\leq n-1, a,b \ne n-2}
    \frac{\hat{s}_{a\;b\;n}}{|a\; b\; n|}\left(y_a- y_b\right) +\alpha \tau
    \frac{\hat{s}_{n-2\;n-1\;n}}{\xi(x_n-x_{n-1})}(y_n-y_{n-1}), \nonumber\\
  E_{n}^{(2)}
  & = -\tau \sum\limits_{1\leq a< b\leq n-1, a,b \ne n-2} \frac{\hat{s}_{a\;b\;n}}
    {|a\; b\; n|}\left(x_a- x_b\right) - \tau\alpha \frac{\hat{s}_{n-2\;n-1\;n}}
    {\xi}\ .
\end{align}
Thus each scattering equation is of ${\cal O}(\tau)$. Therefore in the
soft factor, we have one power of $\tau$ in the numerator from the
measure, four powers of $\tau$ in the denominator from the scattering
equations and two powers of $\tau$ in denominator from the
Parke-Taylor factor.  Thus, this limit contributes at order
$\tau^{-5}$ and hence it is the leading order contribution for the
next to adjacent puncture soft limit. However, we have not found
suitable linear combinations of these equations that would allow us to
independently deform the contours away from the scattering equations.
On the other hand, if we localise the delta function on variables
$\xi$ and $\alpha$, while we can consider any of the scattering
equations above and solve for $\xi$, we find that solving for $\alpha$
leads to an $n-3$ degree polynomial equation. Thus for generic
parametrisation, the computation seems to be analytically intractable.
We hope to return to this in the future.

\section{Next to Next to Adjacent Soft Limits for $k=3$ Amplitudes}
\label{sec:nnasl3}

In this section we will study the double soft limit for $n$-th
puncture and its next to next to adjacent puncture. We consider
punctures labelled by $n-3$ and $n$ to be soft,
\begin{equation}
  \ese{a}{b}{n-3} = \tau\hat{s}_{a\,b\,n-3}, \qquad \ese{a}{b}{n} =
  \tau\hat{s}_{a\,b\,n}, \qquad \ese{a}{n-3}{n} = \tau^2\hat{s}_{a\,n-3\,n},
  \qquad a, b \ne n-3, n.
\end{equation}
The scattering equations for regular solutions, {\em i.e.}, when the
determinants $|a\;b\;n|$ and $|a\;b\;n-3|$ are non-vanishing as
$\tau\to 0$, are,
\begin{align}
  E_a^{(i)}
  = & \sum\limits_{b,c \ne a, n-3, n} \frac{\ese{a}{b}{c}}{\td{a}{b}{c}}
      \frac{\partial}{\partial x_a^{(i)}}\td{a}{b}{c} = 0\ ,\quad
      \forall a\ , \nonumber\\
  E_{n-3}^{(i)}
  = & \tau \sum\limits_{a,b\ne n-3, n}\frac{\hat{s}_{a\,b\,n-3}}
      {\td{a}{b}{n-3}}\frac{\partial}{\partial x_{n-3}^{(i)}}
      \td{a}{b}{n-3}\nonumber\\
      & \hspace{3cm} + \tau^2\sum\limits_{a\ne n-3, n}
      \frac{\hat{s}_{a\,n-3\,n}}{\td{a}{n-3}{n}}\frac{\partial}
      {\partial x_{n-3}^{(i)}}\td{a}{n-3}{n} =0\ , \nonumber\\
  E_n^{(i)}
  = & \tau \sum\limits_{a,b\ne n-3, n}\frac{\hat{s}_{a\,b\,n}}{\td{a}{b}{n}}
      \frac{\partial}{\partial x_n^{(i)}}\td{a}{b}{n}\nonumber\\
      & \hspace{3cm} + \tau^2
      \sum\limits_{a\ne n-3,n}\frac{\hat{s}_{a\,n-3\,n}}{\td{a}{n-3}{n}}
      \frac{\partial}{\partial x_n^{(i)}}\td{a}{n-3}{n}= 0\ .
\end{align}
Hence the soft factor can be written as,
\begin{align}
  \mathtt{S}^{(3)}_{n,n-3}
  = & \int d^2 x_n\int d^2 x_{n-3}\; \delta^{(2)}(E_n)
      \delta^{(2)}(E_{n-3})\Biggl[\frac{\td{n-5}{n-4}{n-2}}{
      \td{n-5}{n-4}{n-3}\td{n-4}{n-3}{n-2}
      } \nonumber\\
   &  \hspace{2.2cm} \times\frac{\td{n-4}{n-2}{n-1}\td{n-2}{n-1}{1}
     \td{n-1}{1}{2}}{\td{n-3}{n-2}{n-1}\td{n-2}{n-1}{n}\td{n-1}{n}{1}
     \td{n}{1}{2}}\Biggr]^2\ .
\end{align}
In this case contributions from non-degenerate solutions dominate over
that of degenerate ones because in the Parke-Taylor factor none of the
determinants simultaneously contain the punctures $n$ and $(n-3)$,
hence it does not lead to any $\mathcal{O}(\tau)$ contribution in the
denominator for the degenerate case. As a result in the degenerate
case, the measure and the scattering equations together give the
scaling which is less than the $\mathcal{O}(\tau^{-4})$ scaling of the
non-degenerate solutions. In this case the soft factor is expressed as,
\begin{align}
  \mathtt{S}^{(3)}_{n,n-3}
  = &\int d^2 x_n\; \delta^{(2)}(E_n)\left[\frac{\td{n-2}{n-1}{1}
      \td{n-1}{1}{2}}{\td{n-2}{n-1}{n}\td{n-1}{n}{1}\td{n}{1}{2}}
      \right]^2\nonumber\\
    & \times \int d^2 x_{n-3}\delta^{(2)}(E_{n-3})\left[\frac{
      \td{n-5}{n-4}{n-2}\td{n-4}{n-2}{n-1}}{\td{n-5}{n-4}{n-3}
      \td{n-4}{n-3}{n-2}\td{n-3}{n-2}{n-1}}\right]^2\nonumber\\
  &  \nonumber\\
  = & \frac{1}{\tau^4}\Bigg\{\frac{1}{\hat{s}_{n-2\; n-1\; n}\
      \hat{s}_{n\;1\;2} }+ \frac{1}{\sum\limits_{\substack{a=2\\a\neq n-3}}^{n-1}
  \hat{s}_{a\;n\;1}}\left(\frac{1}{\hat{s}_{n-1\; n\; 1}} + \frac{1}
  {\hat{s}_{n\;1\;2}}\right)\nonumber\\
    & \hspace{1cm}+ \frac{1}{\sum\limits_{\substack{a=1\\a\neq n-3}}^{n-2}
  \hat{s}_{a\;n-1\;n}} \left(\frac{1}{\hat{s}_{n-1\; n\; 1}} + \frac{1}
  {\hat{s}_{n-2\; n-1 \; n}} \right)\Bigg\}\Bigg[\frac{1}
  {\hat{s}_{n-5\; n-4\; n-3}\ \hat{s}_{n-3\;n-2\;n-1} }\nonumber\\
    &  \hspace{1cm} + \frac{1}{\sum\limits_{\substack{a=1\\a\neq n-3,\;n-4}}^{n-1}
  \hat{s}_{a\;n-4\;n-3}}\left(\frac{1}{\hat{s}_{n-5\; n-4\; n-3}} +
  \frac{1}{\hat{s}_{n-4\;n-3\;n-2}}\right)\nonumber\\
    &  \hspace{1cm} + \frac{1}{\sum\limits_{\substack{a=1\\a\neq n-3,\;n-2}}^{n-1}
  \hat{s}_{a\;n-3\;n-2}} \left(\frac{1}{\hat{s}_{n-4\; n-3\; n-2}} +
  \frac{1}{\hat{s}_{n-3\; n-2 \; n-1}} \right)\Bigg]\nonumber\\
 = & \ \mathtt{S}^{(3)}_n\mathtt{S}^{(3)}_{n-3}\ ,
\end{align}
where $\mathtt{S}^{(3)}_{n}$ and $\mathtt{S}^{(3)}_{n-3}$ are the
single soft factors corresponding to $n$-th and $(n-3)$-th soft
external states respectively. Similar result holds when any of the
external state from the set $\{3,4,\cdots n-3\}$ along with $n$-th
state are taken to be soft,
\begin{equation}
  {\mathtt{S}^{(3)}_{n,a} \ =\ \mathtt{S}^{(3)}_n\mathtt{S}^{(3)}_{a}}\qquad
  \forall\; a\in\{3,4,\cdots n-3\}\ .
\end{equation}
Therefore, in this case double soft factor is the same as the product
of two single soft factors.  This is analogous to the $k=2$
non-adjacent case discussed in \ref{sec:nasl}.

\section{Discussion}
\label{sec:disn}

In this paper, we derived the double soft limit for adjacent soft
external states for arbitrary $k$.  We also generalised our method to
the double soft limit of the next to adjacent and next to next to
adjacent soft external states.  We found that in the simultaneous
double soft limit leading contribution in case of the adjacent soft
external states scales as $\tau^{-3(k-1)}$ in the $\tau\to 0$ limit.
It follows from a simple scaling argument, in the canonical color
ordering, that when the soft labels $i$ and $j$ in an amplitude are
arranged in such a way that $|i-j|\in\{k,k+1,\cdots n-k+1\}$ then the
double soft factor is a product of two single soft factors and hence
it scales as $\tau^{-2(k-1)}$.  For all intermediate separations, {\em
  i.e.}, $1<|i-j|<k$, the scaling exponent is linear in $|i-j|$ for
a cyclic ordering. 

The fact that these soft limits involve higher order poles seems to
indicate that these amplitudes could be relevant for computation of
loop diagrams, see e.g., \cite{Franco:2014csa}\footnote{Relation
  between the $(3,6)$ amplitude and four point one-loop integrand in
  cubic biadjoint scalar field theory is studied in
  \cite{Abhishek:2020sdr}.}.  Alternatively, the higher order poles
could be a signature of composite particles or multiparticle states
contributing to the amplitude.  It would be interesting to check if
this is true by further studying the factorisation in detail.  In a
recent paper by some of the authors\cite{Abhishek:2020sdr}, the
relation of the double soft factor with the cluster algebra was
presented.  It would be interesting to generalise our results to
multiple soft theorem.  There are no efficient techniques to date
that can compute arbitrary $(k,n)$ amplitudes.  However, one may be
able to bootstrap these amplitudes by knowing the structure of
multi-soft factors.  Whether multi-soft factors themselves are
amplitudes of some theory is worth exploring but we will leave it for
the future.

\section*{Acknowledgements}
 We would like to thank Subhroneel Chakrabarti,
Alok Laddha, R. Loganayagam, Biswajit Sahoo, and Ashoke Sen for
discussion.  One of us (APS) would like to thank ICTS for virtual
hospitality and the participants of the seminar for asking interesting
and pertinent questions.  We would like to thank the organisers and
the participants of the Recent Developments in S-matrix Theory program
(ICTS/rdst2020/07) where a preliminary version of this work was presented.

\begin{appendix}

\section{Next to Adjacent Non-degenerate Solutions for $k=3$}
\label{sec:app1}

In this appendix, we will discuss the non-degenerate contributions to
the subleading results for the next to adjacent double soft limit.
For the non-degenerate configuration we neglect terms of
$\mathcal{O}(\tau^{2})$ in the scattering equations
\eq{scatt_eqn_non-adj}.  In the homogeneous coordinates we then
obtain,
\begin{eqnarray}\label{next_to_adj_non-deg}
\mathtt{S}^{(3)}_{\text{non-deg}} &= &\oint\frac{\left(\sigma_{n}\; d^{2}\sigma_{n}\right)\left(X_{1}\;Y_{1}\;\sigma_{n}\right)}{\underline{\sum\limits_{b,c}\frac{s_{n\;b\;c}\left(X_{1}\;b\;c\right)}{\left(\sigma_{n}\;b\;c\right)} \sum\limits_{b,c}\frac{s_{n\;b\;c}\left(Y_{1}\;b\;c\right)}{\left(\sigma_{n}\;b\;c\right)}}} \oint\frac{\left(\sigma_{n-2}\;d^{2}\sigma_{n-2}\right)\left(X_{2}\;Y_{2}\;\sigma_{n-2}\right)}{\underline{\sum\limits_{b,c}\frac{s_{n-2\;b\;c}\left(X_{2}\;b\;c\right)}{\left(\sigma_{n-2}\;b\;c\right)} \sum\limits_{b,c}\frac{s_{n-2\;b\;c}\left(Y_{2}\;b\;c\right)}{\left(\sigma_{n-2}\;b\;c\right)}}} \nonumber\\
                                    && \hspace{-2.2cm} \times \left[\frac{\left(n-4\; n-3\; n-1\right)\left(n-3\; n-1\; 1\right)\left(n-1\; 1\; 2\right)}{\left(n-4\; n-3\; \sigma_{n-2}\right)\left(n-3\; \sigma_{n-2}\; n-1\right)\left(\sigma_{n-2}\; n-1\; \sigma_{n}\right)\left(n-1\; \sigma_{n}\; 1\right)\left(\sigma_{n}\;1\;2\right)}\right]^{2}\ .
\end{eqnarray}
To evaluate this integral we will employ the global residue theorem,
for which we will deform the contour from the poles of the scattering
equations and pick up contributions from the singularities of the
integrand.  First we deform the $\sigma_{n-2}$ contour and encounter
poles at the zeroes of the determinants in the denominator of the
integrand,
\begin{equation}
  \left(n-4\; n-3\; \sigma_{n-2}\right)\ , \qquad \left(n-3\; \sigma_{n-2}\; n-1\right)\ , \qquad \left(\sigma_{n-2}\; n-1\; \sigma_{n}\right)\ .
\end{equation}
While evaluating the integral in \eq{next_to_adj_non-deg}, we
encounter two different kinds of singularities which we will discuss
below by considering those cases one at a time.
\subsection{Collision singularities}
		
In this case $(n-2)$-th puncture collides with any one of the
punctures listed below,
\begin{eqnarray}
  1. \qquad  \sigma_{n-2} & \rightarrow & \sigma_{n-3} \nonumber\\
  2. \qquad   \sigma_{n-2} & \rightarrow & \sigma_{n-1} \nonumber\\
  3. \qquad \sigma_{n-2} & \rightarrow & \sigma_{n-4} \nonumber\\
  4.  \qquad \sigma_{n-2} & \rightarrow & \sigma_{n}
\end{eqnarray}
\subsubsection*{Case 1:
  $\mathbf{\sigma_{n-2}\rightarrow\sigma_{n-3}}$}
		
We choose the following parametrisation,
\begin{equation}
  \sigma_{n-2} = \sigma_{n-3} + \epsilon A.
\end{equation}
where $\epsilon$ is treated as one complex variable, which implies
that the variable $A\in\mathbb{CP}^2$ has only one independent
component.  We choose $X_{2} = \sigma_{n-3}$, and deform the contour
of $\sigma_{n-2}$,
\begin{eqnarray}
  && \oint\frac{\left(n-3\; A\; dA\right)\epsilon^2 d\epsilon\left(n-3\; Y_{2}\; A\right)}{\sum\limits_{b, c\ne n-3}s_{n-2\; b\; c}\sum\limits_{b\ne n-3}\frac{s_{n-2\; n-3\; b}\left(Y_{2}\; b\; n-3\right)}{\epsilon\left(A\; b\; n-3\right)}} \nonumber\\
  &&\hspace{1cm}\times\left[\frac{\left(n-4\; n-3\; n-1\right)}{\underline{\epsilon^{2}\left(n-4\; n-3\; A\right)\left(n-3\; A\; n-1\right)}\left(n-3\; n-1\; \sigma_{n}\right)}\right]^{2} \nonumber\\
  \nonumber\\
  & = & \frac{1}{\sum\limits_{b, c\ne n-3}s_{n-2\; b\; c}} \oint \frac{\left(n-3\; A\; dA\right)\left(n-3\; Y_{2}\; A\right)}{\sum\limits_{b\ne n-3}\frac{s_{n-2\; n-3\; b}\left(Y_{2}\; b\; n-3\right)}{\left(A\; b\; n-3\right)}}\nonumber\\
  \nonumber\\
  &&\hspace{1cm}\times \left[\frac{\left(n-4\; n-3\; n-1\right)}{\underline{\left(n-4\; n-3\; A\right)\left(n-3\; A\; n-1\right)}}\right]^{2} \frac{1}{\left(n-3\; n-1\; \sigma_{n}\right)^{2}} \nonumber\\
  & = & \frac{1}{\sum\limits_{ c\ne n-3}s_{n-2\; n-3\; c}}\left[\frac{1}{s_{n-2\; n-3\; n-4}} + \frac{1}{s_{n-1\; n-2\; n-3}}\right] \frac{1}{\left(n-3\; n-1\; \sigma_{n}\right)^{2}}\ .
\end{eqnarray}
In the last line we reduce the integration to that on $\mathbb{CP}^1$
by treating $n-3$ as a spectator, choose $Y_2= \begin{pmatrix}
  0\\
  1
\end{pmatrix}
$, and $A = \begin{pmatrix}
  1\\
  x_{A}
\end{pmatrix}$.  We are then left with a $\sigma_{n}$ integration to
be performed, however, we notice that it is precisely the single soft
integral,
\begin{eqnarray}
  \mathtt{S}^{(3)}_{1} & = & \frac{1}{\sum\limits_{c\ne n-3}s_{n-2\; n-3\; c}}\left[\frac{1}{s_{n-2\; n-3\; n-4}} + \frac{1}{s_{n-1\; n-2\; n-3}}\right] \nonumber\\
                       && \hspace{1cm}\times \oint\frac{\left(\sigma_{n}d^{2}\sigma_{n}\right)\left(X_{1}Y_{1}\sigma_{n}\right)}{\underline{\sum\limits_{b,c}\frac{s_{nbc}\left(X_{1}bc\right)}{\left(\sigma_{n}bc\right)} \sum\limits_{b,c}\frac{s_{nbc}\left(Y_{1}bc\right)}{\left(\sigma_{n}bc\right)}}} \left[\frac{\left(n-3\; n-1\; 1\right)\left(n-1\; 1\; 2\right)}{\left(n-3\; n-1\; \sigma_{n}\right)\left(n-1\; \sigma_{n}\; 1\right)\left(\sigma_{n}12\right)}\right]^{2}\nonumber\\
                       & = & \left[\frac{1}{\sum\limits_{c\ne n-3}s_{n-2\; n-3\; c}}\left(\frac{1}{s_{n-2\; n-3\; n-4}} + \frac{1}{s_{n-1\; n-2\; n-3}}\right)\right]\nonumber\\
                       && \hspace{1cm}\times \Biggl[ \frac{1}{s_{n-3\; n-1\; n}\;s_{n\; 1\; 2}} + \frac{1}{\sum\limits_{c\ne n-1}s_{n-1\; n\; c}}\left(\frac{1}{s_{n-3\; n-1\; n}} + \frac{1}{s_{n-1\; n\; 1}}\right)\nonumber\\
                       &&\hspace{5cm} + \frac{1}{\sum\limits_{ c\ne 1}s_{n\; 1\; c}}\left(\frac{1}{s_{n\; 1\; 2}} + \frac{1}{s_{n-1\; n\; 1}}\right)\Biggr]\ .
\end{eqnarray}
		
\subsubsection*{Case 2:
  $\mathbf{\sigma_{n-2} \rightarrow \sigma_{n-1}}$}
We parametrize $\sigma_{n-2}$ as,
\begin{equation}
  \sigma_{n-2} = \sigma_{n-1} + \epsilon A\ .
\end{equation}
We choose the reference vector $X_{2} = \sigma_{n-1}$.  The
$\sigma_{n-2}$ integration then becomes,
\begin{eqnarray}\label{def_F(n)}
  && \frac{1}{\sum\limits_{b,c\ne n-1}s_{n-2\; b\; c}} \oint\frac{\left(n-1\; A\; dA\right)\left(n-1\; Y_{2}\; A\right)}{\sum\limits_{c\ne n-1}\frac{s_{n-2\; n-1\; c}\left(Y_{2}\; n-1\; c\right)}{\left(A\; n-1\; c\right)}} \left[\frac{\left(n-3\; n-1\; 1\right)}{\underline{\left(n-3\; A\; n-1\right)\left(A\; n-1\; \sigma_{n}\right)}}\right]^{2} \nonumber\\
  & := & \frac{1}{\sum\limits_{b,c\ne n-1}s_{n-2\; b\; c}} \mathcal{F}\left(\sigma_{n}\right).
\end{eqnarray}
The $\sigma_{n}$ integrations will then be as follows,
\begin{eqnarray}
  \mathtt{S}^{(3)}_{2} &= & \frac{1}{\sum\limits_{b,c\ne n-1}s_{n-2\; b\; c}} \oint\frac{\left(\sigma_{n}\;d^{2}\sigma_{n}\right)\left(X_{1}\;Y_{1}\;\sigma_{n}\right)}{\underline{\sum\limits_{b,c}\frac{s_{n\;b\;c}\left(X_{1}\;b\;c\right)}{\left(\sigma_{n}\;b\;c\right)} \sum\limits_{b,c}\frac{s_{n\;b\;c}\left(Y_{1}\;b\;c\right)}{\left(\sigma_{n}\;b\;c\right)}}} \mathcal{F}\left(\sigma_{n}\right) \nonumber \\[2mm]
                       &&\hspace{5cm}\times\left[\frac{\left(n-1\; 1\; 2\right)}{\left(n-1\; \sigma_{n}\; 1\right)\left(\sigma_{n}\;1\;2\right)}\right]^{2}.
\end{eqnarray}
After contour deformation $\sigma_{n}$ will encounter poles at,
\begin{eqnarray}
  i. \qquad \sigma_{n} & \rightarrow & \sigma_{1} \nonumber\\
  ii. \qquad \sigma_{n} & \rightarrow & \sigma_{n-1} \nonumber\\
  iii. \qquad \sigma_{n} & \rightarrow & \sigma_{2} \nonumber\\
  iv. \qquad \sigma_{n} & \rightarrow & \text{Poles of $\mathcal{F}(\sigma_{n})$}
\end{eqnarray}
		
\paragraph{$i$.  $\sigma_{n} \rightarrow \sigma_{1}$:}
We will choose,
\begin{equation}
  \sigma_{n} = \sigma_{1} + \eta B.
\end{equation}
We set $X_{1} = \sigma_{1}$ to obtain,
\begin{eqnarray}
  \mathtt{S}^{(3)}_{2; 1} & = & \frac{1}{\sum\limits_{b,c\ne n-1}s_{n-2\; b\; c}\sum\limits_{e,f\ne 1}s_{n\;e\;f}}\oint\frac{\left(n-1\; A\; dA\right)\left(n-1\; Y_{2}\; A\right)}{\sum\limits_{c\ne n-1}\frac{s_{n-2\; n-1\; c}\left(Y_{2}\; n-1\; c\right)}{\left(A\; n-1\; c\right)}}  \nonumber\\
                          && \times \left[\frac{\left(n-3\; n-1\; 1\right)}{\underline{\left(n-3\; A\; n-1\right)\left(A\; n-1\; 1\right)}}\right]^{2}\oint\frac{\left(1\; B\; dB\right)\left(1Y_{1}B\right)}{\sum\limits_{c\ne 1}\frac{s_{n\;1\;c}\left(Y_{1}1c\right)}{B\;1\;c}}\left[\frac{\left(n-1\; 1\; 2\right)}{\underline{\left(n-1\; B\; 1\right)\left(B12\right)}}\right]^{2}\nonumber\\
                          & = & \frac{1}{\sum\limits_{b\ne n-1}s_{n-2\; n-1\; b}\sum\limits_{c\ne 1}s_{n\;1\;c}}\left[\frac{1}{s_{n-3\; n-2\; n-1}} + \frac{1}{s_{n-2\; n-1\; 1}}\right]\left[\frac{1}{s_{n-1\; n\; 1}} + \frac{1}{s_{n\; 1\; 2}}\right].\nonumber\\
\end{eqnarray}
\paragraph{$ii$. $\sigma_{n}\rightarrow \sigma_{n-1}$:}
We choose our parametrisation,
\begin{equation}
  \sigma_{n} = \sigma_{n-1} + \eta B\ ,
\end{equation}
with the reference vector $X_{2} = \sigma_{n-1}$. In this case we note
that,
\begin{equation}
  \mathcal{F}\left(\sigma_{n}\right) = \frac{1}{\eta^{2}}\mathcal{F}\left(B\right).
\end{equation}
The $B$ integration then takes the form,
\begin{eqnarray}
  \mathtt{S}^{(3)}_{2;2} & = & \frac{1}{\sum\limits_{b,c\ne n-1}s_{n-2\; b\; c}\sum\limits_{e,f\ne n-1}s_{n\;e\;f}}\nonumber\\
                         &&\hspace{1cm}\times\oint\frac{\left(n-1\; B\; dB\right)\left(n-1\; Y_{2}\; B\right)}{\sum\limits_{c\ne n-1}\frac{s_{n\; n-1\; c}\left(Y_{1}\; n-1\; 1\right)}{\left(B\; n-1\; c\right)}}\Biggl\{\underline{\mathcal{F}\left(B\right)\frac{1}{\left(n-1\; B\; 1\right)^{2}}\Biggr\}}.
\end{eqnarray}
Here $\sigma_{n-1}$ is a spectator and hence both $A$ and $B$
integrals can be converted to $\mathbb{CP}^{1}$ integrals. We choose
$Y_{1} = Y_{2} = \begin{pmatrix}
  0\\
  1
\end{pmatrix}
$, $A = \begin{pmatrix}
  1\\
  x_{A}
\end{pmatrix}$, and $B = \begin{pmatrix}
  1\\
  x_{B}
\end{pmatrix}$.  With this choice the above integrations become,
\begin{eqnarray}
  \mathtt{S}^{(3)}_{2;2} & = & \frac{1}{\sum\limits_{b,c\ne n-1}s_{n-2\; b\; c}\sum\limits_{e,f\ne n-1}s_{n\;e\;f}} \oint\frac{dx_{B}}{\sum\limits_{c\ne n-1}\frac{s_{n\; n-1\; c}}{x_{B}-x_{c}}} \oint\frac{dx_{A}}{\sum\limits_{c\ne n-1}\frac{s_{n-2\; n-1\; c}}{x_{A}-x_{c}}}\nonumber\\
                         &&\hspace{2.5cm}\times\left[\frac{x_{n-3} - x_{1}}{\underline{\left(x_{A}- x_{n-3}\right)\left(x_{A}-x_{B}\right)\left(x_{B}-x_{1}\right)}}\right]^{2}.
\end{eqnarray}
We first evaluate residues at $x_{A} = x_{n-3}, x_{B}$ and then
perform the $x_{B}$ integration. $x_{A}$ has a simple pole at
${x_{n-3}}$ and a double pole at $x_{B}$. Therefore we get,
\begin{eqnarray}
  \mathtt{S}^{(3)}_{2;2} & = & \frac{1}{\sum\limits_{b,c\ne n-1}s_{n-2\; b\; c}\sum\limits_{e,f\ne n-1}s_{n\;e\;f}} \oint\frac{dx_{B}}{\sum\limits_{c\ne n- 1}\frac{s_{n\; n-1\; c}}{x_{B}-x_{c}}} \left[\frac{x_{1} - x_{n-3}}{\underline{\left(x_{B} - x_{n-3}\right)\left(x_{B}- x_{1}\right)}}\right]^{2} \nonumber\\
                         && \times \left[\frac{1}{s_{n-1\; n-2\; n-3}} + \frac{\sum\limits_{c\ne n-1}\frac{s_{n-2\; n-1\; c}}{\left(x_{B}- x_{c}\right)^{2}}}{\left(\sum\limits_{c\ne n-1}\frac{s_{n-2\; n-1\; c}}{x_{B}- x_{c}}\right)^{2}} - \frac{2}{\left(x_{B} - x_{n-3}\right)\sum\limits_{c\ne n-1}\frac{s_{n-2\; n-1\; c}}{x_{B}- x_{c}}}\right]\ .\nonumber\\
\end{eqnarray}
Thus we see $x_{b}$ has simple poles at $x_{n-3}$ and
$x_{1}$. Moreover, the residue at $x_{B} = x_{n-3}$ turns out to
vanish. The last term in the square bracket does not have a pole at
$x_{B} = x_{1}$, hence evaluating $x_{B}$ integral we obtain,
\begin{equation}\label{n->n-1}
  \mathtt{S}^{(3)}_{2;2}  =  \frac{1}{\sum\limits_{b\ne n-1}s_{n-2\; n-1\; b}\sum\limits_{c\ne n-1}s_{n-1\;n\;c}} \left[\frac{1}{s_{n-1\; n\; 1}} \left(\frac{1}{s_{n-3\; n-2\; n-1}} + \frac{1}{s_{n-2\; n-1\; 1}}\right)\right].
\end{equation}
\paragraph{$iii$. $\sigma_{n} \rightarrow \sigma_{2}$:}
We will now parametrize $\sigma_n$ as,
\begin{equation}
  \sigma_{n} = \sigma_{2} + \eta B.
\end{equation}
We find there is no pole as $\eta\to 0$. Hence there is no
contribution from this singularity.
		
\paragraph{$iv$. Poles of $\mathcal{F}\left(\sigma_{n}\right)$:}
By definition of $\mathcal{F}\left(\sigma_{n}\right)$ in
Eq.(\ref{def_F(n)}), and after performing the $A$ integration we
obtain,
\begin{eqnarray}
  \mathcal{F}\left(\sigma_{n}\right)&=& \oint\frac{\left(n-1\; A\; dA\right)\left(n-1\; Y_{2}\; A\right)}{\sum\limits_{c\ne n-1}\frac{s_{n-2\; n-1\; c}\left(Y_{2}\; n-1\; c\right)}{\left(A\; n-1\; c\right)}} \left[\frac{\left(n-3\; n-1\; 1\right)}{\underline{\left(n-3\; A\; n-1\right)\left(A\; n-1\; \sigma_{n}\right)}}\right]^{2} \nonumber\\
                                    &=& -\left[\frac{(n-3\;n-1\;1)}{(n-3\;n-1\;n)}\right]^2\Biggl[\frac{1}{s_{n-3\; n-2\; n-1}} + \frac{\sum\limits_{c\ne n-1}\frac{s_{n-2\; n-1\; c}}{\left(n\;n-1\;c\right)^{2}}}{\left(\sum\limits_{c\ne n-1}\frac{s_{n-2\; n-1\; c}}{\left(n\;n-1\;c\right)}\right)^{2}} \nonumber\\
                                    &&\hspace{5cm}+ \frac{2}{\left(n-3\;n-1\;n\right)\sum\limits_{c\ne n-1}\frac{s_{n-2\; n-1\; c}}{\left(n\;n-1\;c\right)}}\Biggr].
\end{eqnarray}
From the above expression we can see that other than at
$\sigma_{n} \rightarrow \sigma_{n-1}$ and at
$\sigma_{n} \rightarrow \sigma_{n-3}$, which have already been taken
into account in Eq.(\ref{n->n-1}), no other poles of $\sigma_{n}$
appear from $\mathcal{F}\left(\sigma_{n}\right)$.
\subsubsection*{Case 3:
  $\mathbf{\sigma_{n-2} \rightarrow \sigma_{n-4}}$ }
From the symmetry argument $\sigma_{n-2} \rightarrow \sigma_{n-4}$
does not contribute.
\subsubsection*{Case 4: $\mathbf{\sigma_{n-2}\rightarrow \sigma_{n}}$
}
We choose our parametrization,
\begin{equation}
  \sigma_{n-2}=\sigma_{n}+\epsilon A.
\end{equation}
We will choose $X_{2}=\sigma_{n}$ and obtain the
Eq.(\ref{next_to_adj_non-deg}) to be,
\begin{eqnarray}
  \mathtt{S}^{(3)}_{3} &= &\oint\frac{\left(\sigma_{n}\;d^{2}\sigma_{n}\right)\left(X_{1}\;Y_{1}\;\sigma_{n}\right)}{\underline{\sum\limits_{b,c}\frac{s_{n\;b\;c}\left(X_{1}\;b\;c\right)}{\left(\sigma_{n}\;b\;c\right)} \sum\limits_{b,c}\frac{s_{n\;b\;c}\left(Y_{1}\;b\;c\right)}{\left(\sigma_{n}\;b\;c\right)}}} \oint\frac{\left(\sigma_{n}\;A\;dA\right)\epsilon^2 d\epsilon\left(\sigma_{n}\;Y_{2}\;A\right)}{\underline{\sum\limits_{b,c}\frac{s_{n-2\;b\;c}\left(\sigma_{n}\;b\;c\right)}{\left(\sigma_{n}+\epsilon A \;b\;c\right)} \sum\limits_{b,c}\frac{s_{n-2\;b\;c}\left(Y_{2}\;b\;c\right)}{\left(\sigma_{n}+\epsilon A \;b\;c\right)}}} \nonumber\\
                       &&\times \left[\frac{\left(n-4\; n-3\; n-1\right)\left(n-3\; n-1\; 1\right)\left(n-1\; 1\; 2\right)}{\left(n-4\; n-3\; \sigma_{n}\right)\left(n-3\; \sigma_{n}\; n-1\right)\epsilon\left(A\; n-1\; \sigma_{n}\right)\left(n-1\; \sigma_{n}\; 1\right)\left(\sigma_{n}\;1\;2\right)}\right]^{2}\nonumber\\
                       & = & \oint\frac{\left(\sigma_{n}\;d^{2}\sigma_{n}\right)\left(X_{1}\;Y_{1}\;\sigma_{n}\right)}{\underline{\sum\limits_{b,c}\frac{s_{n\;b\;c}\left(X_{1}\;b\;c\right)}{\left(\sigma_{n}\;b\;c\right)} \sum\limits_{b,c}\frac{s_{n\;b\;c}\left(Y_{1}\;b\;c\right)}{\left(\sigma_{n}\;b\;c\right)}}} \nonumber\\
                       &&\times\oint\frac{\left(\sigma_{n}\;A\;dA\right)\epsilon^2 d\epsilon \left(\sigma_{n}\;Y_{2}\;A\right)}{{\sum\limits_{b,c}\frac{s_{n-2\;b\;c}\left(\sigma_{n}\;b\;c\right)}{\left(\sigma_{n}\;b\;c\right)}\left(1-\epsilon \frac{\left(A\;b\;c\right)}{\left(\sigma_{n}\; b\;c\right)}\right) \sum\limits_{b,c}\frac{s_{n-2\;b\;c}\left(Y_{2}\;b\;c\right)}{\left(\sigma_{n}\;b\;c\right)}\left(1-\epsilon \frac{\left(A\;b\;c\right)}{\left(\sigma_{n} \;b\;c\right)}\right)}} \nonumber\\
                       &&\hspace{0mm} \times \left[\frac{\left(n-4\; n-3\; n-1\right)\left(n-3\; n-1\; 1\right)\left(n-1\; 1\; 2\right)}{\epsilon\left(n-4\; n-3\; \sigma_{n}\right)\left(n-3\; \sigma_{n}\; n-1\right)\underline{\left(A\; n-1\; \sigma_{n}\right)}\left(n-1\; \sigma_{n}\; 1\right)\left(\sigma_{n}\;1\;2\right)}\right]^{2}.\nonumber\\
\end{eqnarray}
By momentum conservation
  $\sum\limits_{b,c\neq n}s_{n-2\ b\;c}+\sum\limits_{c\neq n}s_{n-2\;
    n\;c}=0$, therefore we can neglect
  $\sum\limits_{b,c\neq n}s_{n-2\ b\;c}$, which is of
  $\mathcal{O}(\tau^2)$, in comparison with $\mathcal{O}(\tau)$
  terms.  We then obtain,
\begin{eqnarray}
  \mathtt{S}^{(3)}_{3} & = & {-} \oint\frac{\left(\sigma_{n}\;d^{2}\sigma_{n}\right)\left(X_{1}\;Y_{1}\;\sigma_{n}\right)}{\underline{\sum\limits_{b,c}\frac{s_{n\;b\;c}\left(X_{1}\;b\;c\right)}{\left(\sigma_{n}\;b\;c\right)} \sum\limits_{b,c}\frac{s_{n\;b\;c}\left(Y_{1}\;b\;c\right)}{\left(\sigma_{n}\;b\;c\right)}}} \oint\frac{\left(\sigma_{n}\;A\;dA\right)d\epsilon \left(\sigma_{n}\;Y_{2}\;A\right)}{{\epsilon\sum\limits_{b,c}\frac{s_{n-2\;b\;c} \left(A\;b\;c\right)}{\left(\sigma_{n}\; b\;c\right)} \sum\limits_{b,c}\frac{s_{n-2\;b\;c}\left(Y_{2}\;b\;c\right)}{\left(\sigma_{n}\;b\;c\right)}}} \nonumber\\
                       &&\times \left[\frac{\left(n-4\; n-3\; n-1\right)\left(n-3\; n-1\; 1\right)\left(n-1\; 1\; 2\right)}{\left(n-4\; n-3\; \sigma_{n}\right)\left(n-3\; \sigma_{n}\; n-1\right)\underline{\left(A\; n-1\; \sigma_{n}\right)}\left(n-1\; \sigma_{n}\; 1\right)\left(\sigma_{n}\;1\;2\right)}\right]^{2}.\nonumber\\
\end{eqnarray}
Now consider $\sigma_{n}\rightarrow \sigma_{n-1}$ with the
parametrization,
\begin{equation}
  \sigma_{n}=\sigma_{n-1}+\eta B.
\end{equation}
We will choose $X_{1}=\sigma_{n-1}$ and also
$Y_{2}=\sigma_{n-1}$. With this choice we have,
\begin{eqnarray}
  \mathtt{S}^{(3)}_{3} & = & {-} \oint\frac{\eta^{2}d\eta\left(n-1\;B\;dB\right)\left(n-1\;Y_{1}\;B\right)}{{\sum\limits_{b,c\neq n-1}\frac{s_{n\;b\;c}\left(n-1\;b\;c\right)}{\left(n-1\;b\;c\right)} \sum\limits_{c}\frac{s_{n\;n-1\;c}\left(Y_{1}\;n-1\;c\right)}{\eta\left(B\;n-1\;c\right)}}}\nonumber\\
                       &&\times \oint\frac{d\epsilon\left(n-1\;A\;dA\right) \eta\left(B\;n-1\;A\right)}{{\epsilon\sum\limits_{c}\frac{s_{n-2\;n-1\;c} \left(A\;n-1\;c\right)}{\eta\left(B\; n-1\; c\right)} \sum\limits_{b,c\neq n-1}\frac{s_{n-2\;b\;c}\left(n-1\;b\;c\right)}{\left(n-1\;b\;c\right)}}} \nonumber\\
                       && \times \left[\frac{\left(n-4\; n-3\; n-1\right)\left(n-3\; n-1\; 1\right)\left(n-1\; 1\; 2\right)}{\eta^{3}\underline{\left(n-4\; n-3\; n-1\right)\left(n-3\; B\; n-1\right)\left(A\; n-1\; B\right)\left(n-1\; B\; 1\right)\left(n-1\;1\;2\right)}}\right]^{2}\nonumber\\[5mm]
                       & = & {-} \oint\frac{d\eta\left(B\;n-1\;dB\right)\left(Y_{1}\;n-1\;B\right)}{{\eta\sum\limits_{b,c\neq n-1}s_{n\;b\;c} \sum\limits_{c}\frac{s_{n\;n-1\;c}\left(Y_{1}\;n-1\;c\right)}{\left(B\;n-1\;c\right)}}} \oint\frac{d\epsilon\left(A\;n-1\;dA\right) \left(A\;n-1\;B\right)}{{\epsilon\sum\limits_{c}\frac{s_{n-2\;n-1\;c} \left(A\;n-1\;c\right)}{\left(B \;n-1 \;c\right)} \sum\limits_{b,c\neq n-1}s_{n-2\;b\;c}}} \nonumber\\
                       && \hspace{3.5cm} \times\left[\frac{\left(n-3\; n-1\; 1\right)}{\underline{\left(n-3\;\; n-1 B\right)\left(A\; n-1\; B\right)\left( B\;n-1\; 1\right)}}\right]^{2}.
\end{eqnarray}
We can now take $\sigma_{n-1}$ as a spectator, and choose
$Y_{1}=\begin{pmatrix}
  0\\
  1
\end{pmatrix}
$, $A = \begin{pmatrix}
  1\\
  x_{A}
\end{pmatrix}$, and $B = \begin{pmatrix}
  1\\
  x_{B}
\end{pmatrix}$.  Thus we have,
\begin{eqnarray}
  \mathtt{S}^{(3)}_{3} & = & {-}\frac{1}{\sum\limits_{b,c\neq n-1}s_{n\;b\;c}\sum\limits_{b,c\neq n-1}s_{n-2\;b\;c}} \oint\frac{\left(B\;n-1\;dB\right)\left(Y_{1}\;n-1\;B\right)}{{ \sum\limits_{c}\frac{s_{n\;n-1\;c}\left(Y_{1}\;n-1\;c\right)}{\left(B\;n-1\;c\right)}}} \nonumber\\
                       &&\times \oint\frac{\left(A\;n-1\;dA\right) \left(A\;n-1\;B\right)}{{\sum\limits_{c}\frac{s_{n-2\;n-1\;c} \left(A\;n-1\;c\right)}{\left(B\; n-1\; c\right)} }}\left[\frac{\left(n-3\; n-1\; 1\right)}{\underline{\left(n-3\; n-1 \;B\right)\left(A\; n-1\; B\right)\left( B\;n-1\; 1\right)}}\right]^{2}\nonumber\\
                       & = & {-}\frac{1}{\sum\limits_{b,c\neq n-1}s_{n\;b\;c}\sum\limits_{b,c\neq n-1}s_{n-2\;b\;c}} \oint\frac{dx_{B}}{{ \sum\limits_{c}\frac{s_{n\;n-1\;c}}{\left(x_{c}-x_{B}\right)}}} \oint\frac{dx_{A} \left(x_{B}-x_{A}\right)}{{\sum\limits_{c}\frac{s_{n-2\;n-1\;c} \left(x_{c}-x_{A}\right)}{\left(x_{c}-x_{B}\right)} }} \nonumber\\
                       && \hspace{5cm} \times\left[\frac{\left(x_{1}-x_{n-3}\right)}{\underline{\left(x_{B}-x_{n-3}\right)\left(x_{B}-x_{A}\right)\left(x_{1}-x_{B}\right)}}\right]^{2}.
\end{eqnarray}
Taking $x_{A}=x_{B}+\epsilon$ and performing the integrations we get,
\begin{eqnarray}
  \mathtt{S}^{(3)}_{3} & = & \frac{-1}{\left(\sum\limits_{b\neq n-1}s_{n-1\;n\;c}\right)\left(\sum\limits_{c\neq n-1}s_{n-2\;n-1\;c}\right)^2}\left(\frac{1}{s_{n-3\;n-1\;n}}+\frac{1}{s_{n-1\;n\;1}}\right).
\end{eqnarray}
It can be checked, by explicit computations, that the other
possibilities, namely, 
$(i)\;\sigma_{n} \rightarrow \sigma_{1},\;(ii)\;\sigma_{n} \rightarrow
\sigma_{2},\;(iii)\;\sigma_{n} \rightarrow
\sigma_{n-3},\;(iv)\;\sigma_{n} \rightarrow \sigma_{n-4}$, will not
contribute.
%
		
\subsection{Collinear singularities}
		
This singularity lies at the codimension 2 boundary where
$\sigma_{n-2}$ is at the intersection of two lines, one of them
containing $\sigma_{n-4}$ and $\sigma_{n-3}$, and other line
containing $\sigma_{n-1}$ and $\sigma_{n}$.  Let $\xi$ be the point of
intersection of these two lines then,
\begin{equation}
  \left(n-4 \; n-3\; \xi\right) = 0, \qquad \left(\xi\; n-1\; \sigma_{n}\right) = 0.
\end{equation}
We choose the parametrization as,
\begin{equation}
  \sigma_{n-2} = \alpha\sigma_{n-1} + \beta\sigma_{n-3} + \xi.
\end{equation}
The $\sigma_{n-2}$ integration becomes,
\begin{eqnarray}
  && \oint \frac{\left(\xi\; n-1\; n-3\right)d\alpha d\beta }{\left[\frac{s_{n-2\; n-3\; n-4}\left(X_{2}\; n-3\; n-4\right)}{\alpha\left(n-1\; n-3\; n-4\right)} + \sum\limits_{b,c \ne \left(n-3, n-4\right)}\frac{s_{n-2\; b\; c}\left(X_{2}\;b\;c\right)}{\left(\xi\; b\;c\right)}\right]} \nonumber\\
  &&\hspace{0.5cm} \times\frac{\left(X_{2}\; Y_{2} \; \alpha\sigma_{n-1}+\beta\sigma_{n-3}+ \xi\right)}{\left[\frac{s_{n-2\; n-3\; n-4}\left(Y_{2}\; n-3\; n-4\right)}{\alpha\left(n-1\; n-3\; n-4\right)} + \sum\limits_{b,c \ne \left(n-3, n-4\right)}\frac{s_{n-2\; b\; c}\left(Y_{2}\;b\;c\right)}{\left(\xi\; b\;c\right)}\right]} \nonumber \\
  &&\hspace{0.5cm}\times\left[\frac{1}{\underline{\alpha\beta}\left(n-3\; \xi\; n-1\right)\left(n-3\; n-1\; \sigma_{n}\right)}\right]^{2}. 
\end{eqnarray}
We choose the reference vectors as $X_{2} = \sigma_{n-1}$ and
$Y_{2} = \xi$ to obtain,
\begin{equation}
  \frac{1}{s_{n-2\; n-3\; n-4}\sum\limits_{b,c\ne\left(n-3, n-4\right)}s_{n-2\; b\; c}} \left[\frac{1}{\left(n-3\; n-1\; \sigma_{n}\right)}\right]^{2}.
\end{equation}
The $\sigma_{n}$ integration then takes the form,
\begin{eqnarray}
  \mathtt{S}^{(3)}_{\text{col}} & = & \frac{1}{s_{n-2\; n-3\; n-4}\sum\limits_{b,c\ne\left(n-3, n-4\right)}s_{n-2\; b\; c}} \nonumber\\
                                && \times \oint\frac{\left(\sigma_{n}\;d^{2}\sigma_{n}\right)\left(X_{1}\;Y_{1}\;\sigma_{n}\right)}{\underline{\sum\limits_{b,c}\frac{s_{n\;b\;c}\left(X_{1}\;b\;c\right)}{\left(\sigma_{n}\;b\;c\right)} \sum\limits_{b,c}\frac{s_{n\;b\;c}\left(Y_{1}\;b\;c\right)}{\left(\sigma_{n}\;b\;c\right)}}} \left[\frac{\left(n-3\; n-1\; 1\right)\left(n-1\; 1\; 2\right)}{\left(n-3\; n-1\; \sigma_{n}\right)\left(n-1\; \sigma_{n}\; 1\right)\left(\sigma_{n}\;1\;2\right)}\right]^{2}\nonumber\\
                                & = & \left[\frac{1}{s_{n-2\; n-3\; n-4}\sum\limits_{b,c\ne\left(n-3, n-4\right)}s_{n-2\; b\; c}}\right]\nonumber\\
                                && \hspace{0cm}\times \Biggl[ \frac{1}{s_{n-3\; n-1\; n}s_{n\; 1\; 2}} + \frac{1}{\sum\limits_{c\ne n-1}s_{n-1\; n\; c}}\left(\frac{1}{s_{n-3\; n-1\; n}} + \frac{1}{s_{n-1\; n\; 1}}\right) \nonumber\\
                                &&\hspace{4cm}+ \frac{1}{\sum\limits_{c\ne 1}s_{n\; 1\; c}}\left(\frac{1}{s_{n\; 1\; 2}} + \frac{1}{s_{n-1\; n\; 1}}\right)\Biggr].
\end{eqnarray}
Now the double soft factor for the non-degenerate configuration where
$n$-th and $(n-2)$-th external states are going soft is given as,
\begin{align}
  \mathtt{S}^{(3)}_{\text{non-deg}}
  & = \mathtt{S}^{(3)}_{1} + \mathtt{S}^{(3)}_{2;1} + \mathtt{S}^{(3)}_{2;2} + \mathtt{S}^{(3)}_{3} + \mathtt{S}^{(3)}_{\text{col}}\ .
\end{align}
We find that the above soft factor scales as $\mathcal{O}(\tau^{-4})$,
which is subleading compared to the degenerate case.
\end{appendix}

\bibliography{grkn.bib}

\begin{thebibliography}{10}
\providecommand{\url}[1]{\texttt{#1}}
\providecommand{\urlprefix}{URL }
\expandafter\ifx\csname urlstyle\endcsname\relax
  \providecommand{\doi}[1]{doi:\discretionary{}{}{}#1}\else
  \providecommand{\doi}{doi:\discretionary{}{}{}\begingroup
  \urlstyle{rm}\Url}\fi
\providecommand{\eprint}[2][]{\url{#2}}

\bibitem{Parke:1986gb}
S.~J. Parke and T.~Taylor,
\newblock \emph{{An Amplitude for $n$ Gluon Scattering}},
\newblock Phys. Rev. Lett. \textbf{56}, 2459 (1986),
\newblock \doi{10.1103/PhysRevLett.56.2459}.

\bibitem{Nair:1988bq}
V.~Nair,
\newblock \emph{{A Current Algebra for Some Gauge Theory Amplitudes}},
\newblock Phys. Lett. B \textbf{214}, 215 (1988),
\newblock \doi{10.1016/0370-2693(88)91471-2}.

\bibitem{Berends:1989hf}
F.~A. Berends, W.~Giele and H.~Kuijf,
\newblock \emph{{Exact and Approximate Expressions for Multi - Gluon
  Scattering}},
\newblock Nucl. Phys. B \textbf{333}, 120 (1990),
\newblock \doi{10.1016/0550-3213(90)90225-3}.

\bibitem{Witten:2003nn}
E.~Witten,
\newblock \emph{{Perturbative gauge theory as a string theory in twistor
  space}},
\newblock Commun. Math. Phys. \textbf{252}, 189 (2004),
\newblock \doi{10.1007/s00220-004-1187-3},
\newblock \eprint{hep-th/0312171}.

\bibitem{Roiban:2004yf}
R.~Roiban, M.~Spradlin and A.~Volovich,
\newblock \emph{{On the tree level S matrix of Yang-Mills theory}},
\newblock Phys. Rev. D \textbf{70}, 026009 (2004),
\newblock \doi{10.1103/PhysRevD.70.026009},
\newblock \eprint{hep-th/0403190}.

\bibitem{Cachazo:2004kj}
F.~Cachazo, P.~Svrcek and E.~Witten,
\newblock \emph{{MHV vertices and tree amplitudes in gauge theory}},
\newblock JHEP \textbf{09}, 006 (2004),
\newblock \doi{10.1088/1126-6708/2004/09/006},
\newblock \eprint{hep-th/0403047}.

\bibitem{Britto:2005fq}
R.~Britto, F.~Cachazo, B.~Feng and E.~Witten,
\newblock \emph{{Direct proof of tree-level recursion relation in Yang-Mills
  theory}},
\newblock Phys. Rev. Lett. \textbf{94}, 181602 (2005),
\newblock \doi{10.1103/PhysRevLett.94.181602},
\newblock \eprint{hep-th/0501052}.

\bibitem{ArkaniHamed:2008yf}
N.~Arkani-Hamed and J.~Kaplan,
\newblock \emph{{On Tree Amplitudes in Gauge Theory and Gravity}},
\newblock JHEP \textbf{04}, 076 (2008),
\newblock \doi{10.1088/1126-6708/2008/04/076},
\newblock \eprint{0801.2385}.

\bibitem{Arkani-Hamed:2016byb}
N.~Arkani-Hamed, J.~L. Bourjaily, F.~Cachazo, A.~B. Goncharov, A.~Postnikov and
  J.~Trnka,
\newblock \emph{{Grassmannian Geometry of Scattering Amplitudes}},
\newblock Cambridge University Press,
\newblock ISBN 978-1-107-08658-6, 978-1-316-57296-2,
\newblock \doi{10.1017/CBO9781316091548} (2016), \eprint{1212.5605}.

\bibitem{Arkani-Hamed:2017tmz}
N.~Arkani-Hamed, Y.~Bai and T.~Lam,
\newblock \emph{{Positive Geometries and Canonical Forms}},
\newblock JHEP \textbf{11}, 039 (2017),
\newblock \doi{10.1007/JHEP11(2017)039},
\newblock \eprint{1703.04541}.

\bibitem{Arkani-Hamed:2017mur}
N.~Arkani-Hamed, Y.~Bai, S.~He and G.~Yan,
\newblock \emph{{Scattering Forms and the Positive Geometry of Kinematics,
  Color and the Worldsheet}},
\newblock JHEP \textbf{05}, 096 (2018),
\newblock \doi{10.1007/JHEP05(2018)096},
\newblock \eprint{1711.09102}.

\bibitem{He:2018pue}
S.~He, G.~Yan, C.~Zhang and Y.~Zhang,
\newblock \emph{{Scattering Forms, Worldsheet Forms and Amplitudes from
  Subspaces}},
\newblock JHEP \textbf{08}, 040 (2018),
\newblock \doi{10.1007/JHEP08(2018)040},
\newblock \eprint{1803.11302}.

\bibitem{Salvatori:2018aha}
G.~Salvatori,
\newblock \emph{{1-loop Amplitudes from the Halohedron}},
\newblock JHEP \textbf{12}, 074 (2019),
\newblock \doi{10.1007/JHEP12(2019)074},
\newblock \eprint{1806.01842}.

\bibitem{Arkani-Hamed:2018rsk}
N.~Arkani-Hamed, C.~Langer, A.~Yelleshpur~Srikant and J.~Trnka,
\newblock \emph{{Deep Into the Amplituhedron: Amplitude Singularities at All
  Loops and Legs}},
\newblock Phys. Rev. Lett. \textbf{122}(5), 051601 (2019),
\newblock \doi{10.1103/PhysRevLett.122.051601},
\newblock \eprint{1810.08208}.

\bibitem{Banerjee:2018tun}
P.~Banerjee, A.~Laddha and P.~Raman,
\newblock \emph{{Stokes polytopes: the positive geometry for $\phi^{4}$
  interactions}},
\newblock JHEP \textbf{08}, 067 (2019),
\newblock \doi{10.1007/JHEP08(2019)067},
\newblock \eprint{1811.05904}.

\bibitem{Raman:2019utu}
P.~Raman,
\newblock \emph{{The positive geometry for $\phi^{p}$ interactions}},
\newblock JHEP \textbf{10}, 271 (2019),
\newblock \doi{10.1007/JHEP10(2019)271},
\newblock \eprint{1906.02985}.

\bibitem{Aneesh:2019ddi}
P.~B. Aneesh, M.~Jagadale and N.~Kalyanapuram,
\newblock \emph{{Accordiohedra as positive geometries for generic scalar field
  theories}},
\newblock Phys. Rev. D \textbf{100}(10), 106013 (2019),
\newblock \doi{10.1103/PhysRevD.100.106013},
\newblock \eprint{1906.12148}.

\bibitem{Kalyanapuram:2019nnf}
N.~Kalyanapuram,
\newblock \emph{{Stokes Polytopes and Intersection Theory}},
\newblock Phys. Rev. D \textbf{101}(10), 105010 (2020),
\newblock \doi{10.1103/PhysRevD.101.105010},
\newblock \eprint{1910.12195}.

\bibitem{Aneesh:2019cvt}
P.~B. Aneesh, P.~Banerjee, M.~Jagadale, R.~R. John, A.~Laddha and S.~Mahato,
\newblock \emph{{On positive geometries of quartic interactions: Stokes
  polytopes, lower forms on associahedra and world-sheet forms}},
\newblock JHEP \textbf{04}, 149 (2020),
\newblock \doi{10.1007/JHEP04(2020)149},
\newblock \eprint{1911.06008}.

\bibitem{Salvatori:2019phs}
G.~Salvatori and S.~Stanojevic,
\newblock \emph{{Scattering Amplitudes and Simple Canonical Forms for Simple
  Polytopes}}  (2019),
\newblock \eprint{1912.06125}.

\bibitem{Arkani-Hamed:2019vag}
N.~Arkani-Hamed, S.~He, G.~Salvatori and H.~Thomas,
\newblock \emph{{Causal Diamonds, Cluster Polytopes and Scattering Amplitudes}}
   (2019),
\newblock \eprint{1912.12948}.

\bibitem{He:2020ray}
S.~He, L.~Ren and Y.~Zhang,
\newblock \emph{{Notes on polytopes, amplitudes and boundary configurations for
  Grassmannian string integrals}},
\newblock JHEP \textbf{04}, 140 (2020),
\newblock \doi{10.1007/JHEP04(2020)140},
\newblock \eprint{2001.09603}.

\bibitem{Kalyanapuram:2020vil}
N.~Kalyanapuram and R.~G. Jha,
\newblock \emph{{Positive Geometries for all Scalar Theories from Twisted
  Intersection Theory}},
\newblock Phys. Rev. Res. \textbf{2}(3), 033119 (2020),
\newblock \doi{10.1103/PhysRevResearch.2.033119},
\newblock \eprint{2006.15359}.

\bibitem{Jagadale:2020qfa}
M.~Jagadale and A.~Laddha,
\newblock \emph{{On the Positive Geometry of Quartic Interactions III : One
  Loop Integrands from Polytopes}}  (2020),
\newblock \eprint{2007.12145}.

\bibitem{Cachazo:2013iaa}
F.~Cachazo, S.~He and E.~Y. Yuan,
\newblock \emph{{Scattering in Three Dimensions from Rational Maps}},
\newblock JHEP \textbf{10}, 141 (2013),
\newblock \doi{10.1007/JHEP10(2013)141},
\newblock \eprint{1306.2962}.

\bibitem{Cachazo:2013gna}
F.~Cachazo, S.~He and E.~Y. Yuan,
\newblock \emph{{Scattering equations and Kawai-Lewellen-Tye orthogonality}},
\newblock Phys. Rev. D \textbf{90}(6), 065001 (2014),
\newblock \doi{10.1103/PhysRevD.90.065001},
\newblock \eprint{1306.6575}.

\bibitem{Cachazo:2013hca}
F.~Cachazo, S.~He and E.~Y. Yuan,
\newblock \emph{{Scattering of Massless Particles in Arbitrary Dimensions}},
\newblock Phys. Rev. Lett. \textbf{113}(17), 171601 (2014),
\newblock \doi{10.1103/PhysRevLett.113.171601},
\newblock \eprint{1307.2199}.

\bibitem{Cachazo:2013iea}
F.~Cachazo, S.~He and E.~Y. Yuan,
\newblock \emph{{Scattering of Massless Particles: Scalars, Gluons and
  Gravitons}},
\newblock JHEP \textbf{07}, 033 (2014),
\newblock \doi{10.1007/JHEP07(2014)033},
\newblock \eprint{1309.0885}.

\bibitem{He:2015yua}
S.~He and E.~Y. Yuan,
\newblock \emph{{One-loop Scattering Equations and Amplitudes from Forward
  Limit}},
\newblock Phys. Rev. \textbf{D92}(10), 105004 (2015),
\newblock \doi{10.1103/PhysRevD.92.105004},
\newblock \eprint{1508.06027}.

\bibitem{Cachazo:2015aol}
F.~Cachazo, S.~He and E.~Y. Yuan,
\newblock \emph{{One-Loop Corrections from Higher Dimensional Tree
  Amplitudes}},
\newblock JHEP \textbf{08}, 008 (2016),
\newblock \doi{10.1007/JHEP08(2016)008},
\newblock \eprint{1512.05001}.

\bibitem{Mason:2013sva}
L.~Mason and D.~Skinner,
\newblock \emph{{Ambitwistor strings and the scattering equations}},
\newblock JHEP \textbf{07}, 048 (2014),
\newblock \doi{10.1007/JHEP07(2014)048},
\newblock \eprint{1311.2564}.

\bibitem{Geyer:2014fka}
Y.~Geyer, A.~E. Lipstein and L.~J. Mason,
\newblock \emph{{Ambitwistor Strings in Four Dimensions}},
\newblock Phys. Rev. Lett. \textbf{113}(8), 081602 (2014),
\newblock \doi{10.1103/PhysRevLett.113.081602},
\newblock \eprint{1404.6219}.

\bibitem{Casali:2015vta}
E.~Casali, Y.~Geyer, L.~Mason, R.~Monteiro and K.~A. Roehrig,
\newblock \emph{{New Ambitwistor String Theories}},
\newblock JHEP \textbf{11}, 038 (2015),
\newblock \doi{10.1007/JHEP11(2015)038},
\newblock \eprint{1506.08771}.

\bibitem{Geyer:2015bja}
Y.~Geyer, L.~Mason, R.~Monteiro and P.~Tourkine,
\newblock \emph{{Loop Integrands for Scattering Amplitudes from the Riemann
  Sphere}},
\newblock Phys. Rev. Lett. \textbf{115}(12), 121603 (2015),
\newblock \doi{10.1103/PhysRevLett.115.121603},
\newblock \eprint{1507.00321}.

\bibitem{Geyer:2015jch}
Y.~Geyer, L.~Mason, R.~Monteiro and P.~Tourkine,
\newblock \emph{{One-loop amplitudes on the Riemann sphere}},
\newblock JHEP \textbf{03}, 114 (2016),
\newblock \doi{10.1007/JHEP03(2016)114},
\newblock \eprint{1511.06315}.

\bibitem{Geyer:2016wjx}
Y.~Geyer, L.~Mason, R.~Monteiro and P.~Tourkine,
\newblock \emph{{Two-Loop Scattering Amplitudes from the Riemann Sphere}},
\newblock Phys. Rev. D \textbf{94}(12), 125029 (2016),
\newblock \doi{10.1103/PhysRevD.94.125029},
\newblock \eprint{1607.08887}.

\bibitem{Geyer:2017ela}
Y.~Geyer and R.~Monteiro,
\newblock \emph{{Gluons and gravitons at one loop from ambitwistor strings}},
\newblock JHEP \textbf{03}, 068 (2018),
\newblock \doi{10.1007/JHEP03(2018)068},
\newblock \eprint{1711.09923}.

\bibitem{Geyer:2018xwu}
Y.~Geyer and R.~Monteiro,
\newblock \emph{{Two-Loop Scattering Amplitudes from Ambitwistor Strings: from
  Genus Two to the Nodal Riemann Sphere}},
\newblock JHEP \textbf{11}, 008 (2018),
\newblock \doi{10.1007/JHEP11(2018)008},
\newblock \eprint{1805.05344}.

\bibitem{Berkovits:2019bbx}
N.~Berkovits, M.~Guillen and L.~Mason,
\newblock \emph{{Supertwistor description of ambitwistor strings}},
\newblock JHEP \textbf{01}, 020 (2020),
\newblock \doi{10.1007/JHEP01(2020)020},
\newblock \eprint{1908.06899}.

\bibitem{Arkani-Hamed:2019mrd}
N.~Arkani-Hamed, S.~He and T.~Lam,
\newblock \emph{{Stringy Canonical Forms}}  (2019),
\newblock \eprint{1912.08707}.

\bibitem{Arkani-Hamed:2020cig}
N.~Arkani-Hamed, T.~Lam and M.~Spradlin,
\newblock \emph{{Positive configuration space}}  (2020),
\newblock \eprint{2003.03904}.

\bibitem{He:2020onr}
S.~He, Z.~Li, P.~Raman and C.~Zhang,
\newblock \emph{{Stringy canonical forms and binary geometries from
  associahedra, cyclohedra and generalized permutohedra}}  (2020),
\newblock \eprint{2005.07395}.

\bibitem{Cachazo:2018wvl}
F.~Cachazo, N.~Early, A.~Guevara and S.~Mizera,
\newblock \emph{{$\Delta$-algebra and scattering amplitudes}},
\newblock JHEP \textbf{02}, 005 (2019),
\newblock \doi{10.1007/JHEP02(2019)005},
\newblock \eprint{1812.01168}.

\bibitem{Cachazo:2019ngv}
F.~Cachazo, N.~Early, A.~Guevara and S.~Mizera,
\newblock \emph{{Scattering Equations: From Projective Spaces to Tropical
  Grassmannians}},
\newblock JHEP \textbf{06}, 039 (2019),
\newblock \doi{10.1007/JHEP06(2019)039},
\newblock \eprint{1903.08904}.

\bibitem{Franco:2014csa}
S.~Franco, D.~Galloni, A.~Mariotti and J.~Trnka,
\newblock \emph{{Anatomy of the Amplituhedron}},
\newblock JHEP \textbf{03}, 128 (2015),
\newblock \doi{10.1007/JHEP03(2015)128},
\newblock \eprint{1408.3410}.

\bibitem{Borges:2019csl}
F.~Borges and F.~Cachazo,
\newblock \emph{{Generalized Planar Feynman Diagrams: Collections}}  (2019),
\newblock \eprint{1910.10674}.

\bibitem{Cachazo:2019xjx}
F.~Cachazo, A.~Guevara, B.~Umbert and Y.~Zhang,
\newblock \emph{{Planar Matrices and Arrays of Feynman Diagrams}}  (2019),
\newblock \eprint{1912.09422}.

\bibitem{Guevara:2020lek}
A.~Guevara and Y.~Zhang,
\newblock \emph{{Planar Matrices and Arrays of Feynman Diagrams: Poles for
  Higher $k$}}  (2020),
\newblock \eprint{2007.15679}.

\bibitem{Sepulveda:2019vrz}
D.~Garc{\'\i}a~Sep{\'u}lveda and A.~Guevara,
\newblock \emph{{A Soft Theorem for the Tropical Grassmannian}}  (2019),
\newblock \eprint{1909.05291}.

\bibitem{Schwab}
B.~U.~W. Schwab and A.~Volovich,
\newblock \emph{{Subleading Soft Theorem in Arbitrary Dimensions from
  Scattering Equations}},
\newblock Phys. Rev. Lett. \textbf{113}(10), 101601 (2014),
\newblock \doi{10.1103/PhysRevLett.113.101601},
\newblock \eprint{1404.7749}.

\bibitem{Afkhami}
N.~Afkhami-Jeddi,
\newblock \emph{{Soft Graviton Theorem in Arbitrary Dimensions}}  (2014),
\newblock \eprint{1405.3533}.

\bibitem{Zlotnikov}
M.~Zlotnikov,
\newblock \emph{{Sub-sub-leading soft-graviton theorem in arbitrary
  dimension}},
\newblock JHEP \textbf{10}, 148 (2014),
\newblock \doi{10.1007/JHEP10(2014)148},
\newblock \eprint{1407.5936}.

\bibitem{Kalousios}
C.~Kalousios and F.~Rojas,
\newblock \emph{{Next to subleading soft-graviton theorem in arbitrary
  dimensions}},
\newblock JHEP \textbf{01}, 107 (2015),
\newblock \doi{10.1007/JHEP01(2015)107},
\newblock \eprint{1407.5982}.

\bibitem{DoubleSoftPRD}
F.~Cachazo, S.~He and E.~Y. Yuan,
\newblock \emph{{New Double Soft Emission Theorems}},
\newblock Phys. Rev. \textbf{D92}(6), 065030 (2015),
\newblock \doi{10.1103/PhysRevD.92.065030},
\newblock \eprint{1503.04816}.

\bibitem{VolovichZlotnikov}
A.~Volovich, C.~Wen and M.~Zlotnikov,
\newblock \emph{{Double Soft Theorems in Gauge and String Theories}},
\newblock JHEP \textbf{07}, 095 (2015),
\newblock \doi{10.1007/JHEP07(2015)095},
\newblock \eprint{1504.05559}.

\bibitem{Saha:2016kjr}
A.~P. Saha,
\newblock \emph{{Double Soft Theorem for Perturbative Gravity}},
\newblock JHEP \textbf{09}, 165 (2016),
\newblock \doi{10.1007/JHEP09(2016)165},
\newblock \eprint{1607.02700}.

\bibitem{Saha:2017yqi}
A.~P. Saha,
\newblock \emph{{Double soft limit of the graviton amplitude from the
  Cachazo-He-Yuan formalism}},
\newblock Phys. Rev. D \textbf{96}(4), 045002 (2017),
\newblock \doi{10.1103/PhysRevD.96.045002},
\newblock \eprint{1702.02350}.

\bibitem{Chakrabarti:2017zmh}
S.~Chakrabarti, S.~P. Kashyap, B.~Sahoo, A.~Sen and M.~Verma,
\newblock \emph{{Testing Subleading Multiple Soft Graviton Theorem for CHY
  Prescription}},
\newblock JHEP \textbf{01}, 090 (2018),
\newblock \doi{10.1007/JHEP01(2018)090},
\newblock \eprint{1709.07883}.

\bibitem{Cachazo:2016ror}
F.~Cachazo, S.~Mizera and G.~Zhang,
\newblock \emph{{Scattering Equations: Real Solutions and Particles on a
  Line}},
\newblock JHEP \textbf{03}, 151 (2017),
\newblock \doi{10.1007/JHEP03(2017)151},
\newblock \eprint{1609.00008}.

\bibitem{Klose:2015xoa}
T.~Klose, T.~McLoughlin, D.~Nandan, J.~Plefka and G.~Travaglini,
\newblock \emph{{Double-Soft Limits of Gluons and Gravitons}},
\newblock JHEP \textbf{07}, 135 (2015),
\newblock \doi{10.1007/JHEP07(2015)135},
\newblock \eprint{1504.05558}.

\bibitem{Georgiou:2015jfa}
G.~Georgiou,
\newblock \emph{{Multi-soft theorems in Gauge Theory from MHV Diagrams}},
\newblock JHEP \textbf{08}, 128 (2015),
\newblock \doi{10.1007/JHEP08(2015)128},
\newblock \eprint{1505.08130}.

\bibitem{Low:2015ogb}
I.~Low,
\newblock \emph{{Double Soft Theorems and Shift Symmetry in Nonlinear Sigma
  Models}},
\newblock Phys. Rev. D \textbf{93}(4), 045032 (2016),
\newblock \doi{10.1103/PhysRevD.93.045032},
\newblock \eprint{1512.01232}.

\bibitem{Chakrabarti:2017ltl}
S.~Chakrabarti, S.~P. Kashyap, B.~Sahoo, A.~Sen and M.~Verma,
\newblock \emph{{Subleading Soft Theorem for Multiple Soft Gravitons}},
\newblock JHEP \textbf{12}, 150 (2017),
\newblock \doi{10.1007/JHEP12(2017)150},
\newblock \eprint{1707.06803}.

\bibitem{AtulBhatkar:2018kfi}
S.~Atul~Bhatkar and B.~Sahoo,
\newblock \emph{{Subleading Soft Theorem for arbitrary number of external soft
  photons and gravitons}},
\newblock JHEP \textbf{01}, 153 (2019),
\newblock \doi{10.1007/JHEP01(2019)153},
\newblock \eprint{1809.01675}.

\bibitem{Jain:2018fda}
D.~Jain and A.~Rudra,
\newblock \emph{{Leading soft theorem for multiple gravitini}},
\newblock JHEP \textbf{06}, 004 (2019),
\newblock \doi{10.1007/JHEP06(2019)004},
\newblock \eprint{1811.01804}.

\bibitem{Marotta:2020oob}
R.~Marotta and M.~Mojaza,
\newblock \emph{{Double-soft behavior of massless closed strings interacting
  with any number of closed string tachyons}},
\newblock JHEP \textbf{08}, 083 (2020),
\newblock \doi{10.1007/JHEP08(2020)083},
\newblock \eprint{2005.05877}.

\bibitem{Cachazo:2019ble}
F.~Cachazo, B.~Umbert and Y.~Zhang,
\newblock \emph{{Singular Solutions in Soft Limits}},
\newblock JHEP \textbf{05}, 148 (2020),
\newblock \doi{10.1007/JHEP05(2020)148},
\newblock \eprint{1911.02594}.

\bibitem{Griffiths:433962}
P.~A. Griffiths and J.~Harris,
\newblock \emph{{Principles of algebraic geometry}},
\newblock Wiley classics library. Wiley, New York, NY,
\newblock \doi{10.1002/9781118032527} (1994).

\bibitem{Abhishek:2020sdr}
M.~Abhishek, S.~Hegde and A.~P. Saha,
\newblock \emph{{One-loop integrand from generalised scattering equations}}
  (2020), \eprint{2012.10916}.

\end{thebibliography}

\nolinenumbers

\end{document}